# Machine learning-assisted surrogate construction for full-core fuel performance analysis

Yifeng Che[1], Joseph Yurko[2], Koroush Shirvan[1]*
[1]*Department of Nuclear Science and Engineering, Massachusetts Institute of Technology, Cambridge, Massachusetts, 02139, USA*
[2]*School of Computing and Information (SCI), University of Pittsburgh, Pennsylvania, 15260, USA*
**Email:* kshirvan@mit.edu

## ABSTRACT

Accurately predicting the behavior of a nuclear reactor requires multiphysics simulation of coupled neutronics, thermal-hydraulics and fuel thermo-mechanics. The fuel thermo-mechanical response provides essential information for operational limits and safety analysis. Traditionally, fuel performance analysis is performed standalone, with spatial-temporal power distribution and thermal boundary conditions calculated from the coupled neutronics-thermal-hydraulics simulation used as input. Such limited one-way coupling is result of cost induced by the full-core fuel performance analysis, which provides more realistic and accurate prediction of the core-wide response than the “peak rod” analysis. The computational burden for full-core fuel performance simulation falls within 12 hours in serial for NRC-licensed code FRAPCON, while becomes unbearable for high fidelity codes like BISON. Therefore, it is desirable to improve the computational efficiency of full-core fuel performance modeling by constructing fast-running surrogate. As such we can utilize fuel performance modeling in the core reload design optimization and improve neutron economy. This work thereby presents methodologies for full-core surrogate construction based on several realistic equilibrium PWR core designs. As a fast and conventional approach, look-up tables (LUTs) are only effective for certain fuel performance quantities of interest (QoIs). Several representative machine-learning (ML) algorithms are therefore introduced to capture the complicated physics for other fuel performance QoIs. Rule-based model is useful as a feature extraction technique to account for the spatial-temporal complexity of the operating conditions. Constructed surrogates achieve at least ten thousand time acceleration compared to FRAPCON with satisfying prediction accuracy. Current work lays foundation for tighter coupling of the fuel performance analysis into the core design optimization framework. It also sets stage for full-core fuel performance analysis with higher fidelity tools like BISON where the computational cost becomes more burdensome.

**Keywords:** Full-core fuel performance analysis; Surrogate modeling; Machine Learning; Core design optimization;

**Nomenclature**

| | |
|---|---|
| CASL | Consortium for Advanced Simulation of Light Water Reactors |
| CDI | cumulative damage index |
| DOE | design of experiments |
| GP | gaussian process regression |
| IFBA | integral fuel burnable absorber |
| LHGR | linear heat generation rate |
| LUT | look-up table |
| LWR | light water reactor |
| MAE | mean absolute error |
| ML | machine learning |
| MPS | missing pellet surface |
| MSE | mean squared error |
| NN | (feed-forward) neural network |
| PCI | pellet-cladding interaction |
| PCMI | pellet-cladding mechanical interaction |
| PLS | partial least-square regression |
| QoI | quantity of interest |
| RBF | radial basis function |
| RF | random forest |
| RL | reinforcement learning |
| RMSE | root mean squared error |
| rRMSE | relative root-mean-square error |
| SCC | stress-corrosion cracking |
| TD | theoretical density |
| XGB | extreme gradient boosted tree |

## 1. Introduction

Nuclear reactor core design involves multiphysics simulations to meet the targeted safety standards, enhance the operational flexibility, and optimize the neutron economy hence fuel consumption. Realistic modeling of a reactor core requires coupling of a complete set of physical phenomena that impacts neutronics, thermal-hydraulics, and fuel thermo-mechanical behaviors at the engineering scale. Neutronics investigates the change in fuel composition and the spatial and temporal distributions of the fission power. Thermal-hydraulics considers the fluid mechanics of the coolant and core-wide heat transfer. Fuel performance analysis depicts the thermo-mechanical behavior of the fuel rods and assemblies as well as chemical processes that take place at the fuel pin level. These three fundamental physics are interdependent in the way that each one provides boundary conditions or essential feedbacks to another. Neutronics calculates the amount of gamma heating to the coolant, as well as power distribution that serves as input to the fuel performance analysis. Thermal-hydraulics estimates the fluid temperature and density which feeds back to neutronics, and the coolant heat transfer coefficients serve as boundary condition in fuel performance analysis. Fuel performance modeling provides essential temperature and density feedback from the fuel rod to both neutronics and thermal-hydraulics.

Simulation capabilities of coupled core-level neutronics and thermal-hydraulics have reached a relatively mature stage with fast runtimes. Such coupled simulations are often augmented with simplified fuel performance feedback including average fuel temperature and thermal expansion of the solid structures (e.g. fuel/clad). Coupling of the fuel performance analysis, on the contrary, remains predominantly a one-way procedure. Output from the coupled neutronics-thermal-hydraulics modeling is normally used as input to the fuel performance modeling, and independent fuel performance calculations are performed for several limiting fuel pins. These limiting pins are typically selected by taking the pins with highest linear heat generation rate and burnup levels from the core-level reactor physics calculations. Recently, there has been more concentration on incorporating the fuel performance analysis into the integrated multiphysics simulations of light water reactors (LWR), as demonstrated in the Virtual Environment for Reactor Applications (VERA) deployed by Consortium for Advanced Simulation of Light Water Reactors (CASL) program [1] [2]. In addition to providing feedbacks to neutronics and thermal-hydraulics during reactor design, fuel performance analysis gives insight to potential failures risks of fuel rods and offers guidance on operational procedures. Thermal-mechanical fuel rod constraints must not be violated during the entire reactor operation (and the dry storage stage for the spent fuel) to maintain the fuel integrity hence radioactivity. Accurate prediction of rod failure risk ensures sufficient safety margin for operation limits, reserves flexibility for operation, and enhances the neutron economy by allowing more aggressive power ramp rates.

Since only fuel performance of small fraction of the rods are typically analyzed, conservative factors are adopted to ensure that these peak rods are representative of the true margin to performance targets. The conservative peak-rod analyses can be undesirable for two reasons. First, the failure risks are typically overestimated, posing unnecessarily stringent requirement on the operation limits. Second, although less time-consuming, the peak-rod analyses do not necessarily reproduce the most hazardous fuel behavior. Some of the limiting fuel performance metrics are not linearly dependent on the power level, but proportional to the integration of power (burnup) or possibly depend on more complicated format of power, particularly during load following. The full-core fuel performance analysis is therefore strongly motivated to more accurately account for the core-wide response.

Despite the fact that the total number of independent code runs can be reduced by exploiting the core symmetry and shuffling scheme, the full-core simulation still involves thousands of rod simulations hence time-consuming. The approximate total runtime for a specific core loading pattern typically falls onto half a day in serial with fuel performance analysis tools FRAPCON. With higher fidelity fuel performance tools like BISON, the total runtime can be as high as 10,000 to 30,000 core-hours on a high computing cluster for a single full-core analysis [2]. In a typical core reload pattern optimization process, currently nuclear utilities use expert input and/or automated tools to evaluate 100s to 1,000,000s (with use of surrogates) of patterns. The objective function for selection of the best patterns is void of any detailed fuel performance for obvious issue due to high computational cost. This motivates creation of fast running accurate surrogate models for fuel performance analysis to achieve a tighter coupling with current core simulators. Particularly, we plan to integrate the surrogate model in a recent physics-informed reinforcement learning (RL) optimization framework [3] [4]. The current constraints and objectives employed in this physics-informed RL framework accommodates only coupled neutronics-thermal-hydraulics response from a commercial licensed code. The use of RL holds promise in significantly reducing the number of needed core pattern evaluations compared to the traditional optimization methods employed by nuclear industry (e.g. simulated annealing or genetic algorithm). Therefore, we plan to leverage the time savings by incorporating more physics to the RL framework to achieve more robust core reload patterns. However, the RL optimization process still will requires tens of thousands of possible core pattern evaluations as the runtime with even low-fidelity fuel performance tools becomes prohibitive. Therefore, surrogate construction is again motivated to significantly reduce the computational cost of the full-core simulation without much sacrifice of prediction accuracy.

The current work investigates useful techniques to construct fast-running surrogates for full-core fuel performance simulation with emphasis on two pragmatically meaningful tasks. The first effort is made to accurately predict the full-core behavior based on a reduced number of independent rod simulations for certain (known) core patterns. The second intention focuses on providing an efficient physics-informed full-core surrogate for copious unobserved core patterns. Runtime of such full-core surrogate should fall onto the timescale of seconds (at least a thousand times faster) to be capable of assisting the developed RL framework by coupling the fuel performance feedback into the objective functions and constraints.

Two categories of methodologies have been explored for surrogate construction: look-up tables (LUTs) as a quick empirical linear mapping, and machine-learning (ML) algorithms as alternative data-driven tools. LUTs, although extensively applied in industry for accurate prediction of average fuel temperature, prove inadequate in predicting some critical fuel performance quantities of interest (QoIs) that embed great nonlinearity. ML algorithms are therefore employed to capture the complicated nonlinear physical phenomena. Feature engineering during data preparation is the key to assure the success of ML algorithms, especially with the complicated spatial-temporal operating history. Rule-based model is investigated in this work for feature extraction and dimensionality reduction. Experimental design is equally important to ensure the appropriateness of the training data such that the entire data range is sufficiently covered by the training samples. For essential fuel performance QoIs, different types of supervised learning algorithms are applied and compared, including linear regression (partial least square regression, PLS), ensembling techniques (random forests, RF, and extreme gradient boosted trees, XGB), neural networks (NN) and gaussian processes regression (GP).

This paper is therefore organized in the following order. Section 2 identifies a complete set of essential fuel performance QoIs that translate to safety design limits during reactor operation. Section 3 presents the full-core fuel performance analysis result for one specific fuel loading pattern. Section 4 illustrates LUT as an empirical surrogate, revealing its suitability and inadequacy for different fuel performance QoIs. Section 5 introduces the ML algorithms along with the feature engineering and experimental design techniques used for data preparation. Section 6 compares the predictive performance of LUTs and ML full-core surrogates under different circumstances: restricted applicability on certain core patterns versus extended applicability to numerous unobserved fuel loading patterns. The desirability of the surrogates requires combined consideration of predictive accuracy and time complexity. Section 7 summarizes the major observations and how the current work sets stage for future works.

## 2. Essential QoIs in full-core fuel performance analysis

Fuel performance analysis simulates the thermo-mechanical behaviors of fuel rods under irradiation, subject to vital constraints for structural integrity of fuel rods. Constraints are placed on the following QoIs based on the potential fuel failure mechanisms in LWRs:

1) **Fuel temperature**. The fuel temperature involves feedbacks from multiple procedures. The temperature field promotes the thermal expansion of fuel pellets, causing decreased gap between the fuel and cladding, which in turn leads to lower fuel temperature. High temperature hence high temperature gradient in the fuel drives the diffusion of gaseous fission products, leading to more release of fission gas into the plenum, deteriorating the gap thermal conductivity and increasing the plenum pressure, which further elevates the fuel temperature. In addition, fuel swelling is promoted at high temperature to accommodate the solid and gaseous fission products as the burnup accumulates, which accelerates the gap closure hence pellet-cladding contact. Moreover, the maximum fuel temperature should never approach the fuel melting point (2,865 ˚C) during normal operation. For these reasons, the maximum fuel temperature is selected as one figure of merit in the current work.

2) **Plenum pressure**. The fuel rod internal pressure rises initially due to elevated gas average temperature at reactor startup and reduced plenum volume caused by fuel expansion and cladding creep down, later due to release of fission gases to the rod void volume. Plenum pressure can also be a strong function of the initial fill gas pressure and presence of annular pellets at top and bottom of the fuel stack that provide additional free volume. In addition, a fraction of Westinghouse PWR rods are coated with integral fuel burnable absorber (IFBA) in PWR fuel assemblies. $^{10}B$ in the IFBA coating burns out quickly in the first power cycle and its transmutation produces helium, contributing significantly to the rod internal pressure. Plenum pressure greater than the coolant pressure can cause tensile cladding stress hence cladding creep-out (outward deformation of the cladding tube). Once the cladding creep-out rate exceeds the fuel swelling rate, the pellet-cladding gap increases, reducing the gap conductivity, consequently increasing the fuel temperature. Increased fuel temperature leads to additional fission gas release, higher rod internal pressure, hence further cladding creep-out and larger gap. To prevent such adverse thermal feedback, the plenum pressure should not rise above the coolant pressure to also avoid the cladding lift-off issue.

3) **Cladding oxidation**. Waterside oxidation of the Zircaloy cladding progresses steadily during normal operation and aggressively under accident scenarios in LWRs. Growth of the oxide layer (brittle ceramic $ZrO_2$) leads to not only clad thinning, but more importantly loss of clad ductility due to enhanced solubility of oxygen in the $\beta$-Zr phase at temperatures above 1200 ˚C. For such reason, the maximum oxidation layer thickness is another key parameter that should not exceed 17% of the as-fabricated clad thickness during accident conditions. The pre-accident oxidation thickness is therefore important to capture since it controls the margin to the accident limit.

4) **Cladding hydrogen pickup**. Absorption and diffusion of hydrogen into the zircaloy cladding occur in parallel with the corrosion process on the outer surface of the cladding. Hydrogen precipitates out as brittle hydride phase in zirconium-based alloys due to relatively low solubility at LWR operation temperatures (80 to 100 ppm) [5]. Uniformly distributed hydride causes gross embrittlement in the cladding, affecting its mechanical behavior. Localized hydride phase forming at the tip of a crack can result in crack extension (delayed hydride cracking) which has been a concern for the fuel structural integrity during transients at sufficiently high burnup as well as dry storage of the spent fuel. For accident conditions, NRC regulates that the total amount of hydrogen generated during reactor operation should not exceed 1% of the hydrogen generated hypothetically if all the cladding material (excluding the plenum) reactors with water [5].

5) **Pellet-cladding interaction (PCI)**. The CASL program identified cladding failure through PCI as one challenge that remains to be properly addressed [6]. Potential PCI failure places conservative restrictions on the operational power ramp rates as PCI failures typically occur during severe local power ramps following low operating power. Pure mechanical interaction (pellet-cladding mechanical interaction, PCMI) and stress corrosion cracking (SCC) are the two contributors to the PCI-induced failure.

a) PCMI challenges the cladding integrity with high stress that possibly tears the cladding apart. Among the three principal stresses in the cladding tube (thin-shell), the axial (or longitudinal) and radial stress are usually less than the hoop stress, which acts circumferential and perpendicular to the axis and radius of the cylinder wall. Therefore, the cladding hoop stress is usually investigated rather than the other stress components. The tensile hoop stress/strain can be viewed as a transformed identifier for occurrence of PCMI. Early in life the gap stays open, and the cladding deforms inward due to combined impact of irradiation creep and net compressive pressure (rod internal pressure lower than the coolant pressure). The cladding hoop stress therefore remains compressive (negative) during this stage. Later in life the pellet-cladding contacts due to fuel thermal expansion and cladding shrinkage, the fuel pellets impose outward force onto the cladding, leading to tensile (positive) hoop stress. Cladding is able to withstand much higher compression than tension due to the asymmetric yield strength in Zirconium alloys [7]. The maximum compressive hoop stress that cladding undergoes due to pressure difference typically falls onto 50-70 MPa in LWRs, far below the compressive yield strength of Zircaloy. The maximum tensile hoop stress due to PCMI, on the contrary, can go up to as high as hundreds of MPa, posing risk to clad rupture. It is also worth mentioning that gap closure hence PCMI are usually delayed (or even do not occur) in IFBA rods compared to regular rods, with the cladding hoop stress being compressive for most of the time. Consequently, IFBA rods are generally less vulnerable to PCMI failure. In conclusion, accurate prediction of tensile hoop stress/strain matters more than the compressive component, especially for non-IFBA rods. For such reason, the maximum hoop stress and hoop strain are included into the essential QoIs as indicators for PCMI.

b) Stress corrosion cracking (SCC) induced by PCMI and corrosive fission products is another principal failure mechanism. There are two mechanisms behind the SCC failures: PCI-SCC, where the pellet radial crack causes stress concentration at the cladding inner surface hence SCC failure; PCI-MPS, where SCC failure happens due to manufacturing-induced missing pellet surfaces (MPS) [8]. Cumulative damage index (CDI) model has been developed to compute the accumulated damage in the cladding, which translates into rod failure probability [9]. Failure risks are calculated based on CDI values for both PCI-SCC and PCI-MPS in the current work based on [10] [11]. Details about the failure risk computation is described in **Appendix A**. Rods with failure risk higher than 0.5 are identified as PCI-vulnerable in the full-core PCI monitoring.

This work thereby considers the following eight QoIs as a complete set of essential fuel performance responses for operational safety: (1) maximum fuel temperature, (2) maximum plenum pressure, (3) maximum oxidation thickness in the cladding, (4) maximum hydrogen concentration in the cladding, (5) maximum hoop stress in the cladding, (6) maximum hoop strain in the cladding, (7) rod failure risk due to PCI-SCC, and (8) rod failure risk due to PCI-MPS.

## 3. Full-core fuel performance analysis

A full-core fuel performance analysis for one specific fuel loading pattern is shown in this section. The full-core simulation is performed using NRC-licensed fuel performance analysis code FRAPCON 4.0, based on the output from CASMO-SIMULATE simulation. CASMO-SIMULATE provides core design parameters and fuel specifications (Table 3.1) as well as spatial-temporal distribution of power in the core, which serve as input to FRAPCON for fuel performance modeling. This work adopts equilibrium 15×15 PWR cores with a quarter symmetry for demonstration of methodologies. Figure 1 shows one representative core loading pattern in a quarter core outputted by the mentioned RL framework. The exact reactor physics characteristics of the core performance is not relevant to this work and not included. However, as it will be seen the core exhibits the typical power and burnup distribution of a 4-Loop Westinghouse PWR plant. Each assembly contains 17×17 fuel rod array, with the number of IFBA rods in each assembly specified at the bottom. The core symmetry and shuffling scheme are exploited to reduce the total number of independent code runs [12], such that only 6864 independent FRAPCON runs are required to model the entire core which contains a total number of 50,592 fuel rods. A shutdown period of 15 days follows the end of each fuel cycle. At the beginning of each cycle, a simplified stepwise history is assumed for power ramp up within a total time of 6 days. The linear heat generation rate (LHGR) rises linearly from zero to 30% full power level in the first 10 hours and holds for 50 hours, subsequently elevates to 80% of full power in the next 30 hours and remains there for 20 hours, then arrives at 90% power level in another 10 hours and stays for 10 hours, and finally reaches the 100% power level in the succeeding 14 hours. Two other distinguished cores that were randomly produced by the optimization algorithm are employed in this work as a representative of the typical core design constraints.

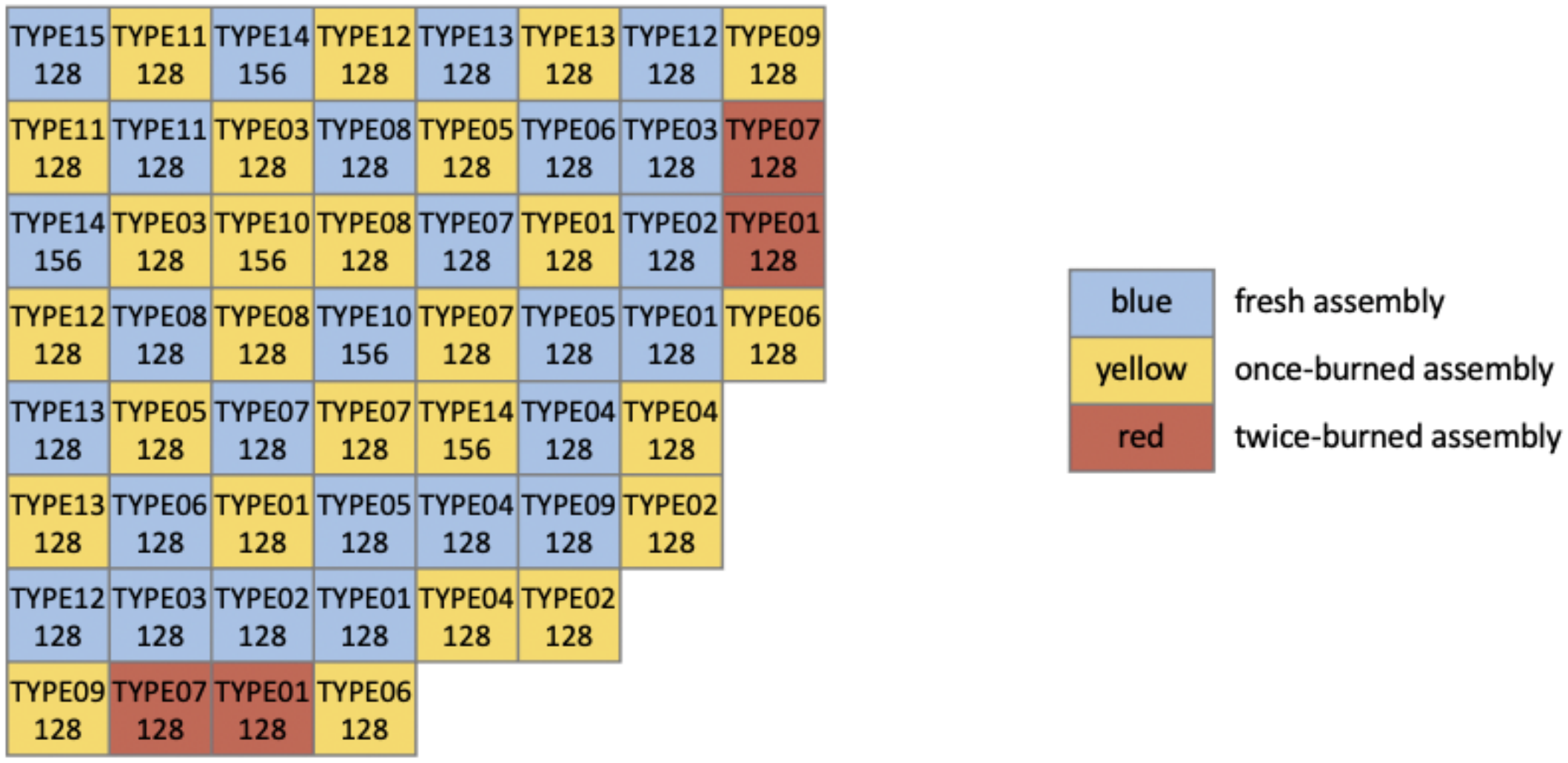

Figure 1. Fuel loading pattern of a quarter core. Blue, yellow and red represent fresh, once-burned and twice-burned assemblies respectively. Number at the bottom denotes the number of IFBA rods in each assembly.

Table 3.1. Core design parameters and fuel characteristics (note: the values are not identical to any of the commercial Westinghouse fuel products)

| **Parameter** | **Value** |
|---|---|
| Fuel/cladding material | $UO_2$ / Zirlo |
| Core thermal power | 3645 MWth |
| Total number of fuel rods | 50,592 |
| Core average linear heat generation rate (LHGR) | 19.70 kW/m |
| Coolant mass flux | 3434.50 kg/($m^2$ s) |
| Coolant pressure | 15.51 MPa |
| Coolant inlet temperature | 565.7 K |
| Active fuel stack length | 365.76 cm |
| Rod outer diameter | 9.50 mm |
| Cladding thickness | 0.5715 mm |
| Fuel-cladding diametric gap thickness | 82.55 µm |
| Annular pellet inner diameter (IFBA rods only) | 3.9172 mm |
| Total length of the drilled section (IFBA rods only) | 40 cm |
| Fuel density | 95.7 % theoretical density (TD) |
| Fuel enrichment | 4.8 % |
| Initial filled pressure for non-IFBA rods | 2.41 MPa |
| Initial filled pressure for IFBA rods | 0.7 MPa |
| Plenum length | 17.5 cm |
| Thickness of $ZrB_2$ coating on pellets (IFBA rods only) | 7.0 µm |
| Boron-10 enrichment in $ZrB_2$ coating | 37 % |
| Density of $ZrB_2$ coating | 90 %TD |

The full core distribution of multiple QoIs are presented in Figure 2. Figure 2(a) demonstrates the distribution of peak linear power over the entire in-reactor life for each rod. Figure 2(b) shows the maximum rod-average burnup, verified between the fuel performance simulation and neutronics calculation. The maximum fuel temperature of the whole core in Figure 2(c) is roughly 1775 K. The highest plenum pressure is up to 15.34 MPa marginally below the coolant pressure (15.51 MPa) as shown in Figure 2(d). Oxidation of the cladding is described in Figure 2(e), where the maximum oxide layer thickness reaches 43 µm, equivalently 7.5% of the total cladding thickness. The core-wide distribution of clad hydrogen concentration in Figure 2(f) is identical to that of the oxidation thickness, as both are physical description of the same waterside corrosion process. Figure 2(g) and Figure 2(h) illustrate the maximum hoop stress and maximum hoop strain respectively. The cladding experiences significant tensile stress/strain during the power ramp up following each shutdown period due to PCI, posing threat to structural integrity. The cladding damage can be estimated by the cumulative damage index (CDI) model [13] [10] for PCI-SCC and PCI-MPS respectively in Figure 2(i) and Figure 2(j). Calculated CDIs correspond to rod failure probabilities (**Appendix A**) as explained in Figure 2(k) and Figure 2(l).

Figure 2(m) to (o) shows core-wide maximum fuel temperature, plenum pressure and maximum cladding hoop stress as a function of time for all rods in the reactor. Red squares mark the fuel pins with peak linear heat generation rate (LHGR), green diamond markers represent the most burnt pins, and the rest of the fuel pins in the reactor are denoted with minor black dots. Among these three QoIs, the most limiting responses occur with neither peak rods nor the most burnt rods. This further proves the necessity to perform full-core analysis for accurate assessment of the safety margin, rather than the “peak rod” analysis which exaggerates the failure risk with conservative projections. It is worth pointing out that vast difference exhibits between regular rods and IFBA rods. Release of helium due to burnout of IFBA coating ($ZrB_2$) adds to the plenum pressure, impacts the gap conductance, delays the gap closure, and leads to a series of changes in corresponding mechanical responses. For such reason, two clear trends present in Figure 2 (m) (n) (o).

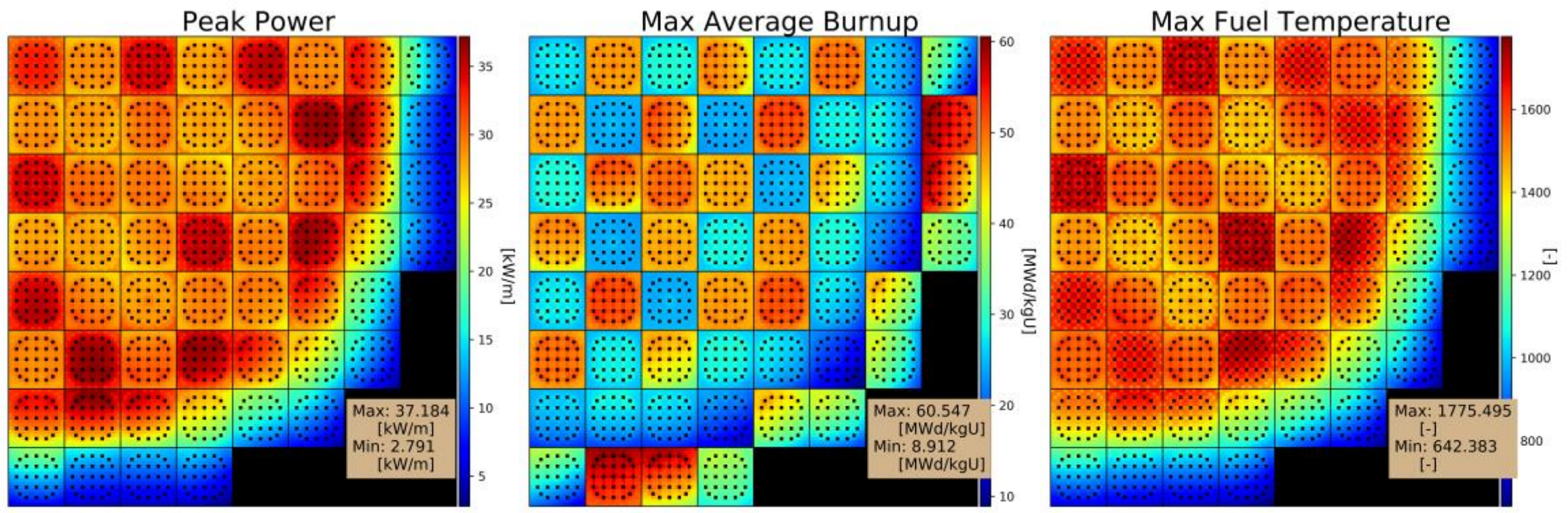

(a) (b) (c)

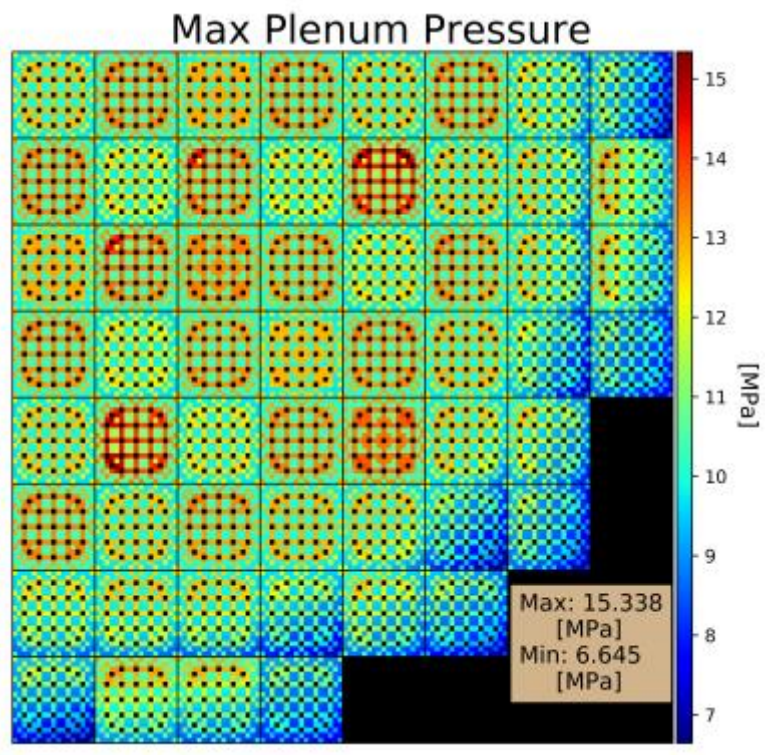

(d)

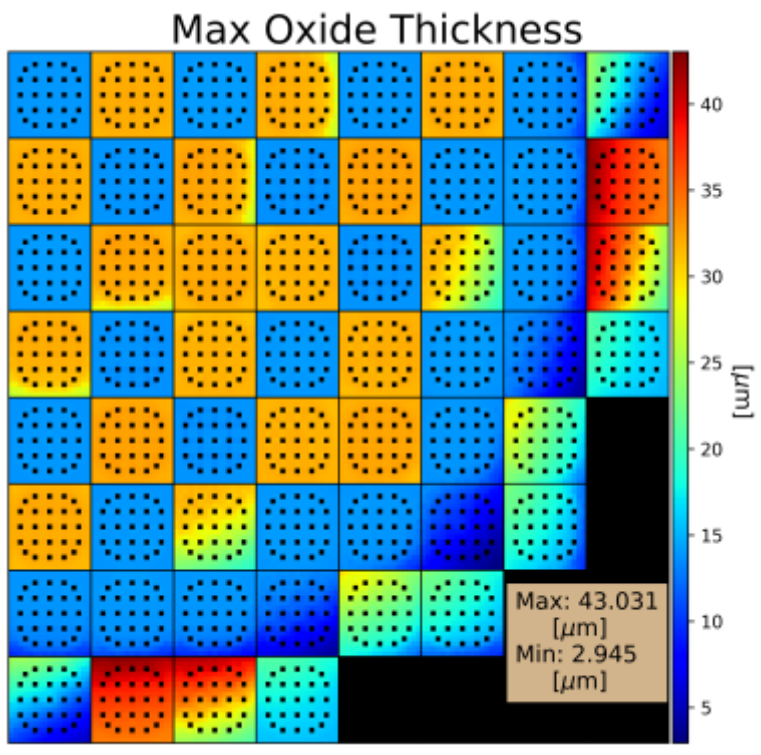

(e)

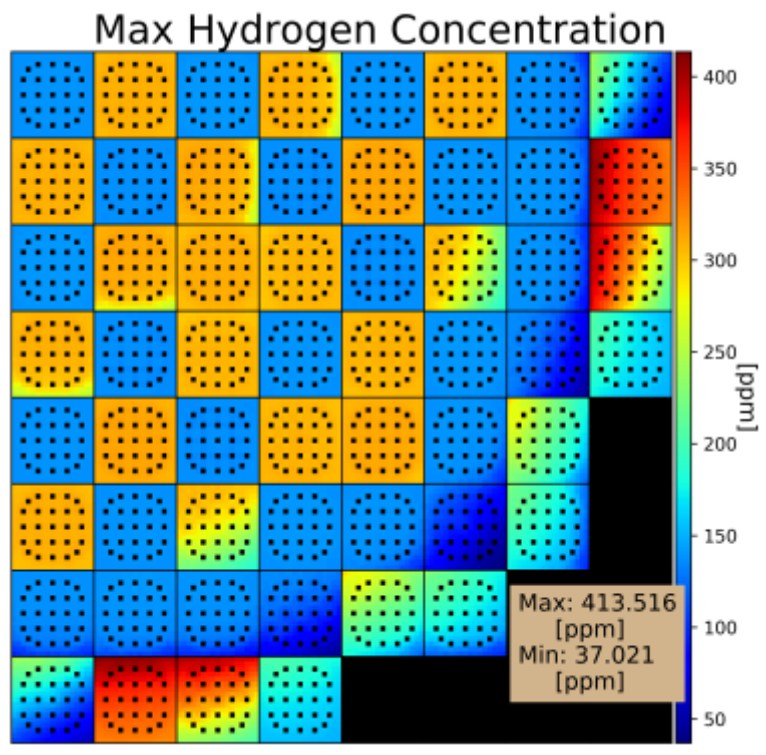

(f)

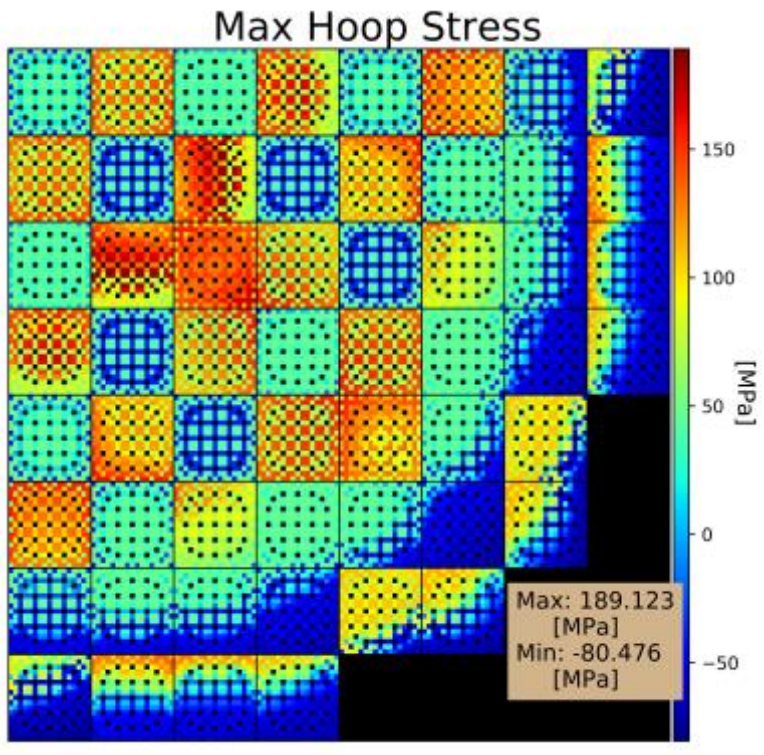

(g)

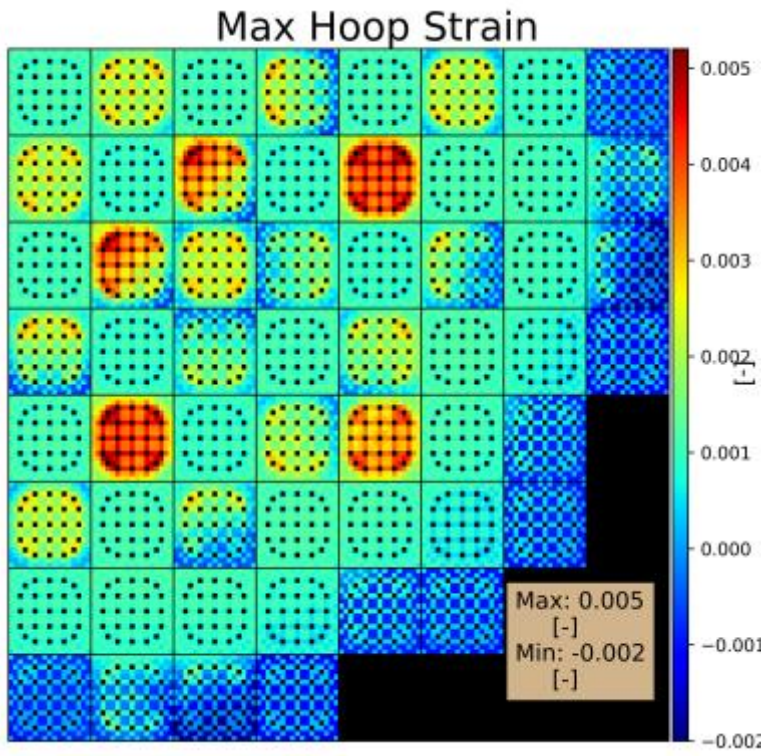

(h)

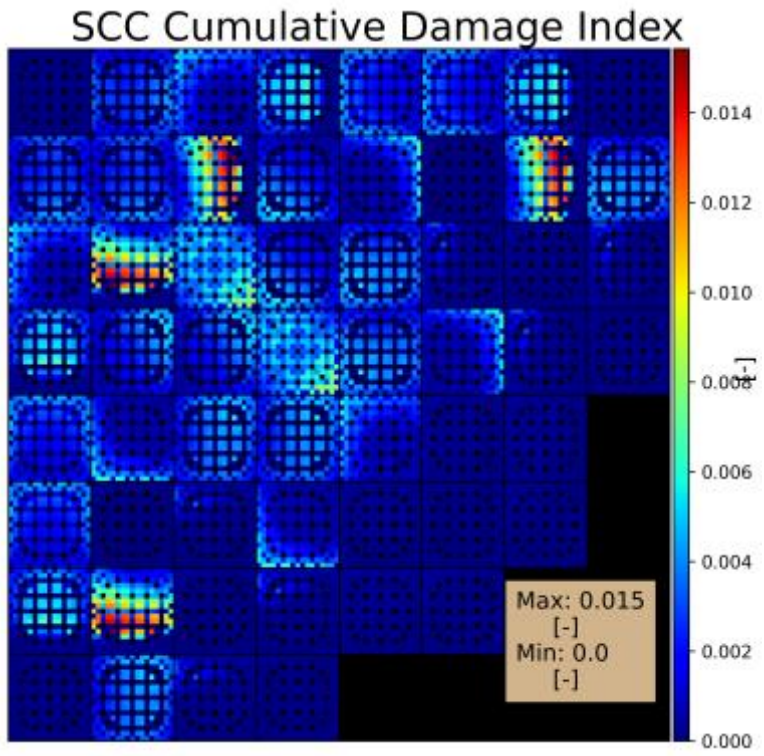

(i)

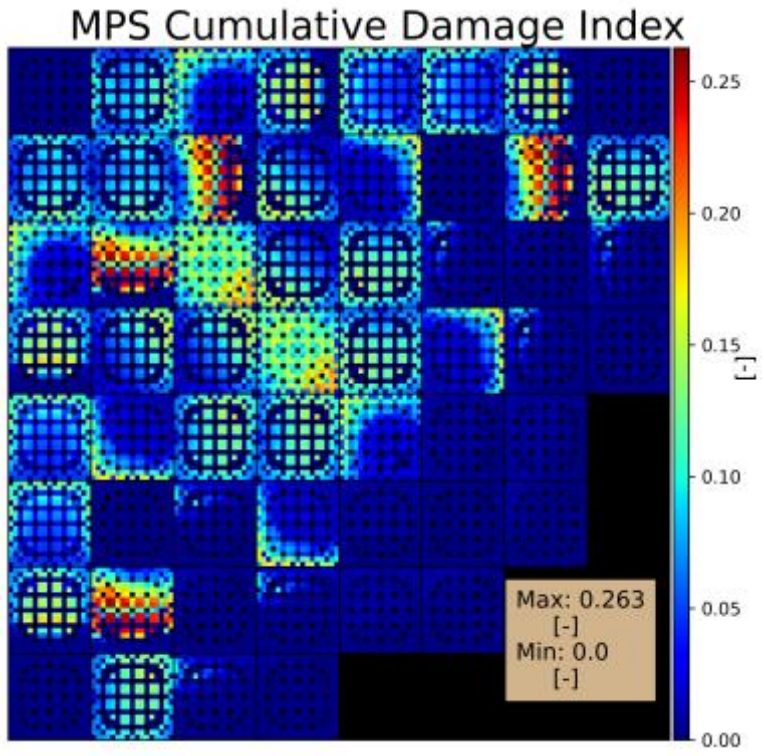

(j)

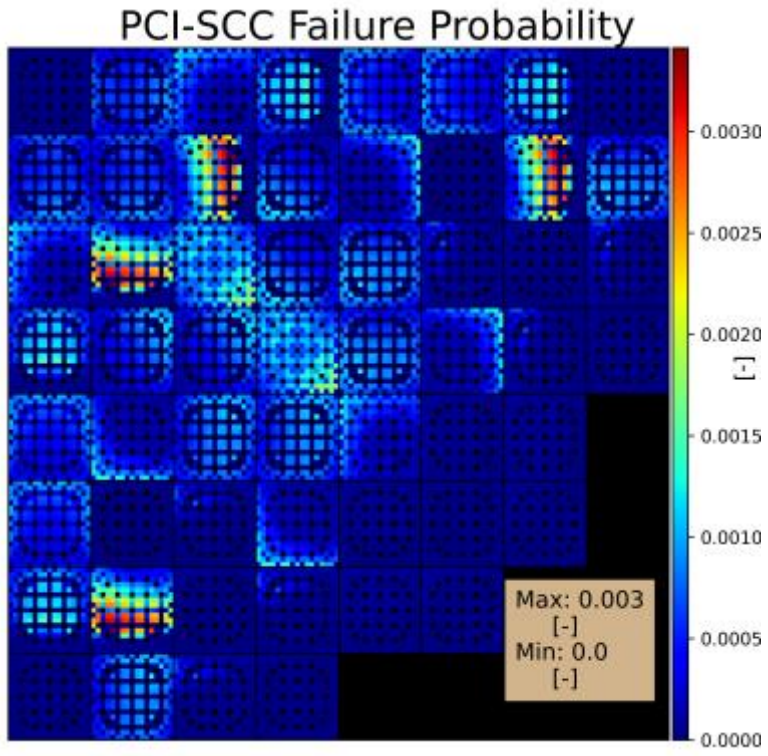

(k)

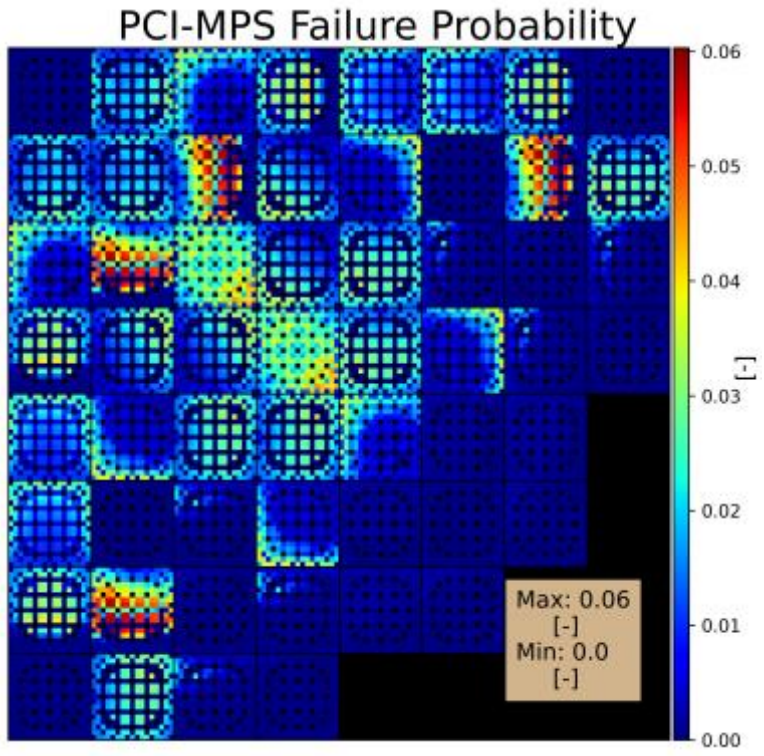

(l)

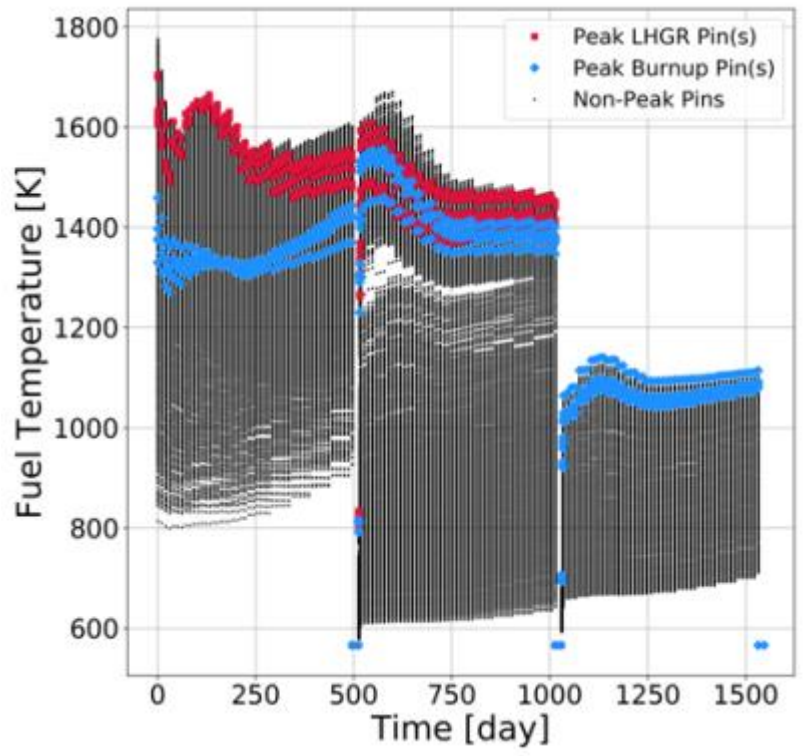

(m)

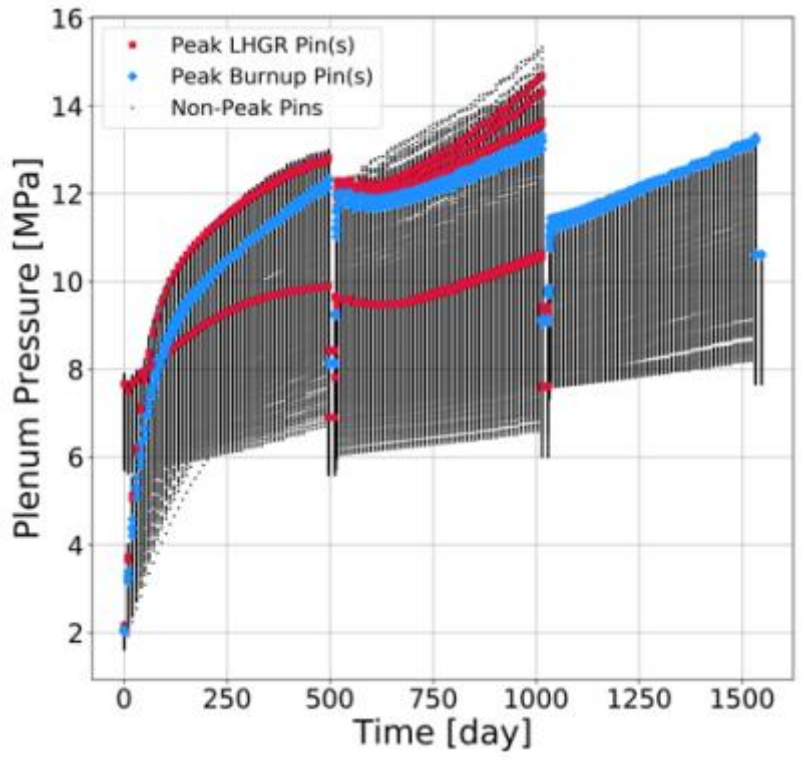

(n)

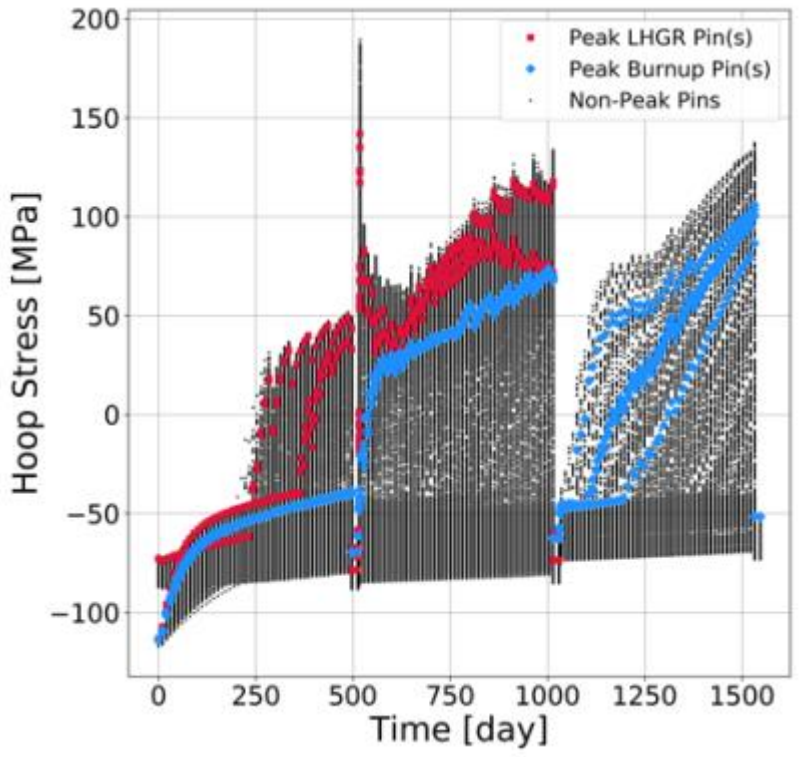

(o)

Figure 2. Full-core fuel performance modeling result for core16. Subfigures (a) to (i) exhibits the core-wide distribution of the maximum values of quantities of interest during the entire reactor operation. Subplots (m) to (o) show the evolution of maximum fuel temperature, plenum pressure and maximum hoop stress as a function of time. Fuel pins with peak linear power are marked with red squares, and the green diamonds mark the most burnt fuel pins.

## 4. Look-up table surrogate

Look-up tables (LUTs) have extensive application in nuclear engineering community as an alternative to physics-based models for material properties, heat transfer, flow regimes and etc. On one hand, the data-driven LUTs are easy to implement and interpret. They may outperform the physics-based models in both prediction and efficiency, especially when the understanding of the underlying physics is imperfect. On the other hand, LUTs are highly empirical. They are typically designed within certain ranges hence fail to extrapolate beyond the ranges of validity.

In fuel performance analysis specifically, LUTs have been widely constructed for average fuel temperature prediction which is coupled to the neutronics analysis for temperature feedback. Such LUTs supply a two-dimensional mapping of the average fuel temperature values at discrete LHGR and burnup levels. The average fuel temperature at one specific LHGR and burnup can be easily accessed by two-dimensional interpolation of the LUT. Validity and accuracy of this LUT arise from the fact that the average fuel temperature calculation solely depends on LHGR as well as integral of local linear power with respect to core height and time (in other words, rod average burnup). The power history that variates with time has little impact onto the average fuel temperature due to the mathematical integration. For this reason, the LUT can be conveniently constructed by running independent FRAPCON simulations at different but constant LHGR levels. Practically, CASMO-SIMULATE simulates the core-wide power and burnup distribution, which can be directly utilized to calculate the corresponding average fuel temperature during numerical iteration.

Despite the desirable performance with average fuel temperature, the suitability of LUTs does not necessarily extend to other fuel performance metrics mentioned in Section 2 as shown in in Figure 3. LUT predictions are plotted versus the true values, along with regions within mean absolute error (MAE) and root mean squared error (RMSE). Predictivity coefficient ($R^2$) compares the variation in the residuals (difference between prediction and true values) to that of the observations. The model is considered more accurate with $R^2$ closer to 1.0, and practically models with $R^2 \geq 0.7$ are accepted as satisfactory. Within this work's interest of investigation, LUTs perform the best for the maximum fuel temperature and the maximum plenum pressure. Forecasts for maximum hydrogen concentration and maximum oxide thickness are less accurate but still acceptable. Estimations for the essential PCI-related QoIs, on the contrary, are poor and require improvement.

Failure of LUTs originates from the complex physical phenomena these QoIs depend on. In fact, few QoI entirely rely on LHGR and average burnup as the average fuel temperature and the maximum fuel temperature do, but are more likely related to the time-dependent operating power. For instance, the maximum cladding hoop stress that serves as an indicator for PCI greatly depends on the occurrence of gap closure, which is closely related with the time-varying power "history" rather than one specific LHGR level. Such history-dependency comprises coupled thermal-mechanical responses much more complicated than simple integration of the spatial-temporal power, i.e., average fuel burnup. Although the power histories are usually distributed within certain finite ranges, in reality a reactor operation can follow countless possible operating histories. LUTs are inherently not designed for such tasks where the number of table entries are infinite.

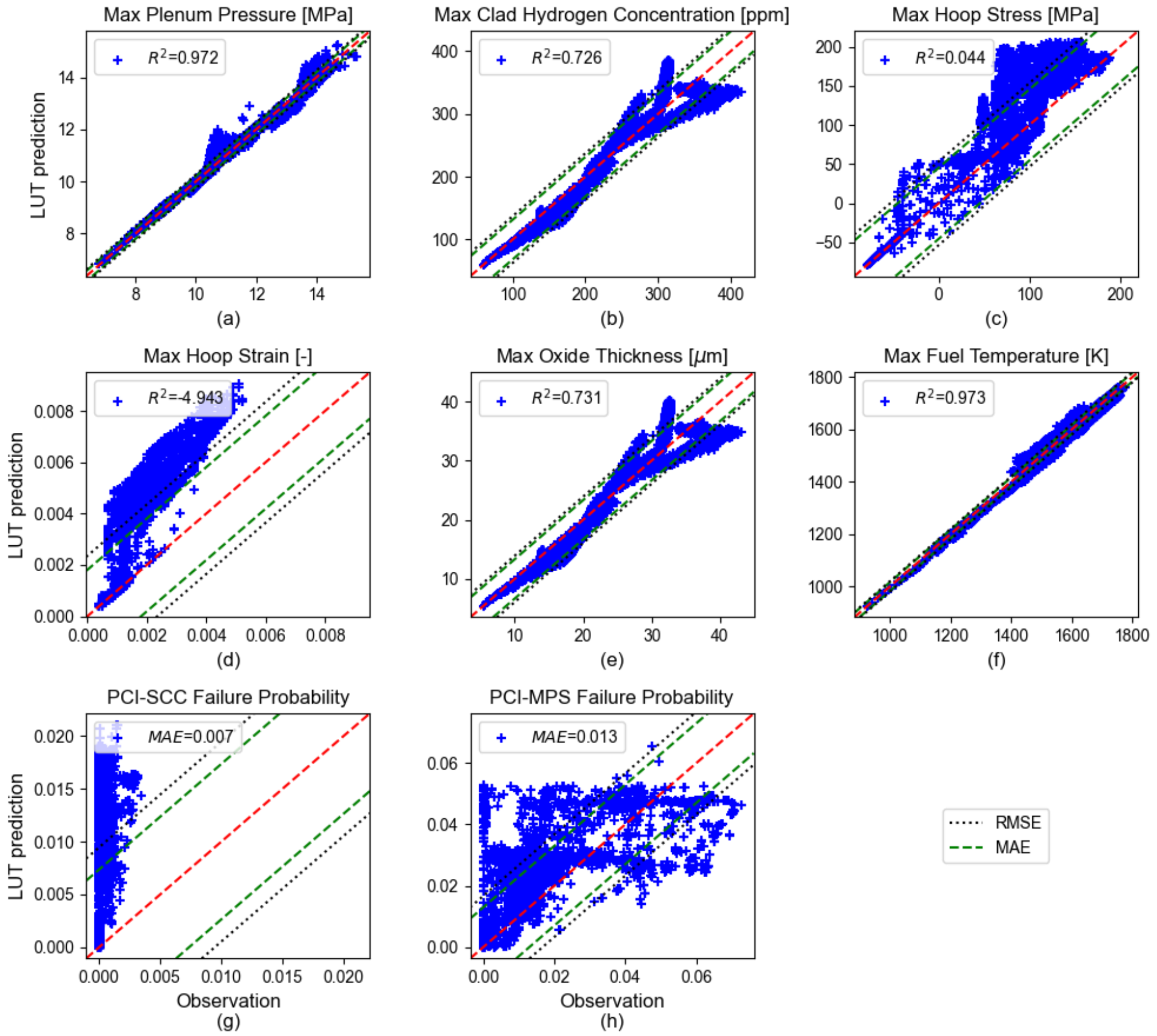

Figure 3. Lookup table prediction of the fuel performance QoIs.

# 5. Machine learning-assisted surrogate

## 5.1 Raw dataset

The reactor physics simulation by CASMO-SIMULATE generates core-wide spatial-temporal power history, which is used as FRAPCON input. The power history decomposes into two components:

(1). Linear heat generation rate (LHGR, averaged over core height) that variates with time, i.e., univariate time series that describe the cross-sectional power distribution.

(2). Axial peaking factors that describe the axial power shape compared to LHGR at each timestep, i.e., multivariate time series that account for local power variations along the core height.

The current work investigates a 3-batch reactor core, where fuel assemblies undergo either one, two or three fuel cycles. Only the analyses for two-cycle rods are presented in this paper as a demonstration of methodology and brevity. Fuel pins in different locations of the same core typically experience different power histories. Figure 4(a) plots the LHGR histories for all fuel pins that go through two fuel cycles in the same core as shown in section 3. Figure 4(b) shows the multivariate time series for axial peaking factor profiles. Each fuel rod is considered as one sample for ML that explores the relationship between spatial-temporal power history and fuel performance QoIs.

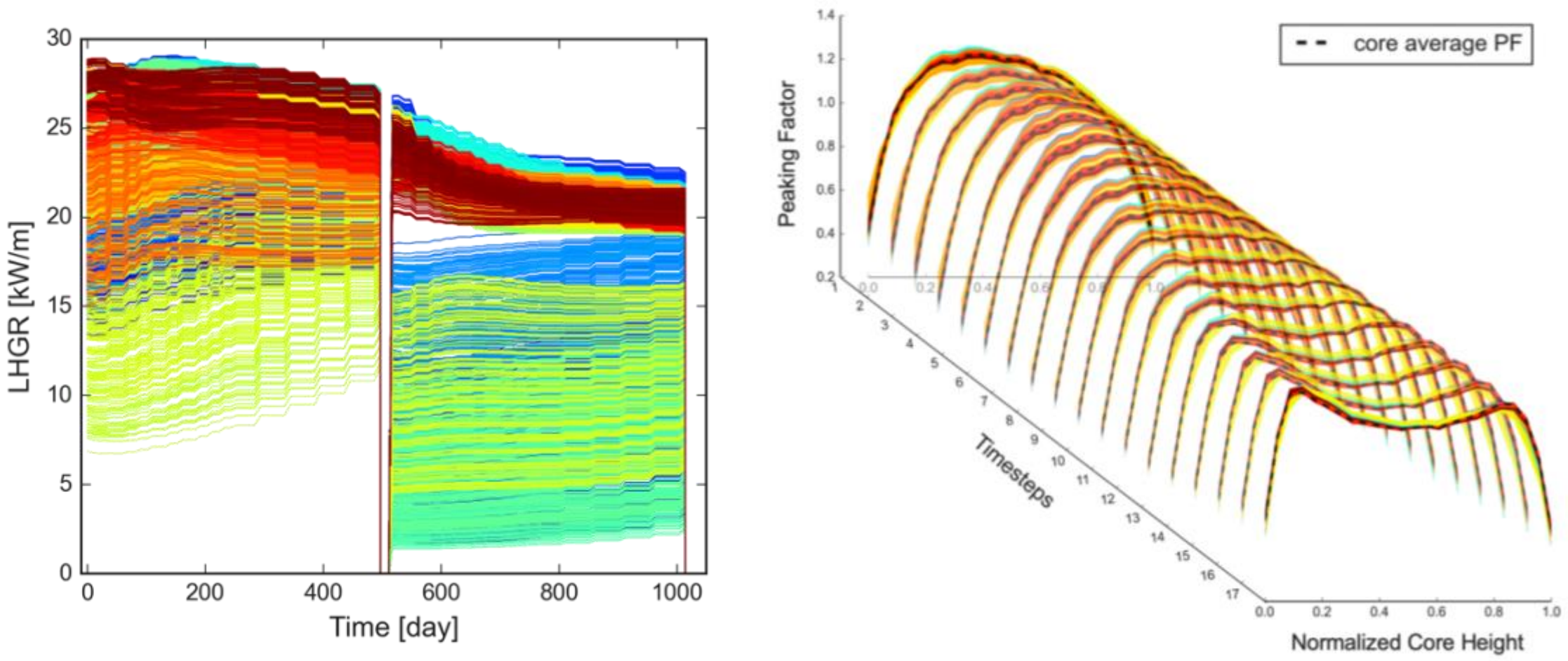

(a) (b)

Figure 4. Power histories for all pins that go through 2 fuel cycles in the same core. (a) LHGR time-series. (b) evolution of axial peaking factor (PF) profiles with time. The black dashed lines denote the core-average PF profile at each time step. Fuel rods are distinguished by color.

## 5.2 Feature engineering: rule-based model

There are multiple ways to perform feature extraction on the same physics problem. Appropriate raw features are those that not only correlates well with the output (good representation of the input space), but also with moderate dimensionality. This work illustrates rule-based modeling as an effective approach to deal with the spatial-temporal complexity of power histories.

The rule-based model representation uses a set of rules to indirectly specify a mathematical model. When the rule-set is highly simpler than the original model it implies, such representation can be especially effective. This work utilizes polynomial regression for purpose of dimensionality reduction. The multivariate time series of PFs is simplified into a univariate time series by taking the maximum axial PF at each timestep to preserve the most limiting fuel performance QoIs. The axial PF profiles are of similar shapes to the core-average PF at each timestep (as shown in in Figure 4(b)), such that the maximum PF value become a valid representative of the axial profile. The two univariate time series (LHGR and maximum PF) are subsequently fitted by 4$^{th}$-order polynomial regression for each power cycle (reactor shutdown and power startup periods removed). Figure 5 shows a schematic of the fitted polynomials for one power cycle of a specific rod. The corresponding polynomial coefficients are considered as representative features of the power history. Additionally, rod type is encoded as a categorical variable with one-hot encoding for IFBA rods and non-IFBA rods. The feature space obtained in this way is thereby 21-dimensional for two-cycle rods. It is also worth mentioning that the errors of the LUT predictions can be treated as an additional feature to boost the performance of ML algorithms when appropriate, given that LUTs utilize a completely different set of features. In this work, the LUT prediction is used to augment the ML performance for the maximum hoop stress, such that the input dimension becomes 22 for this QoI.

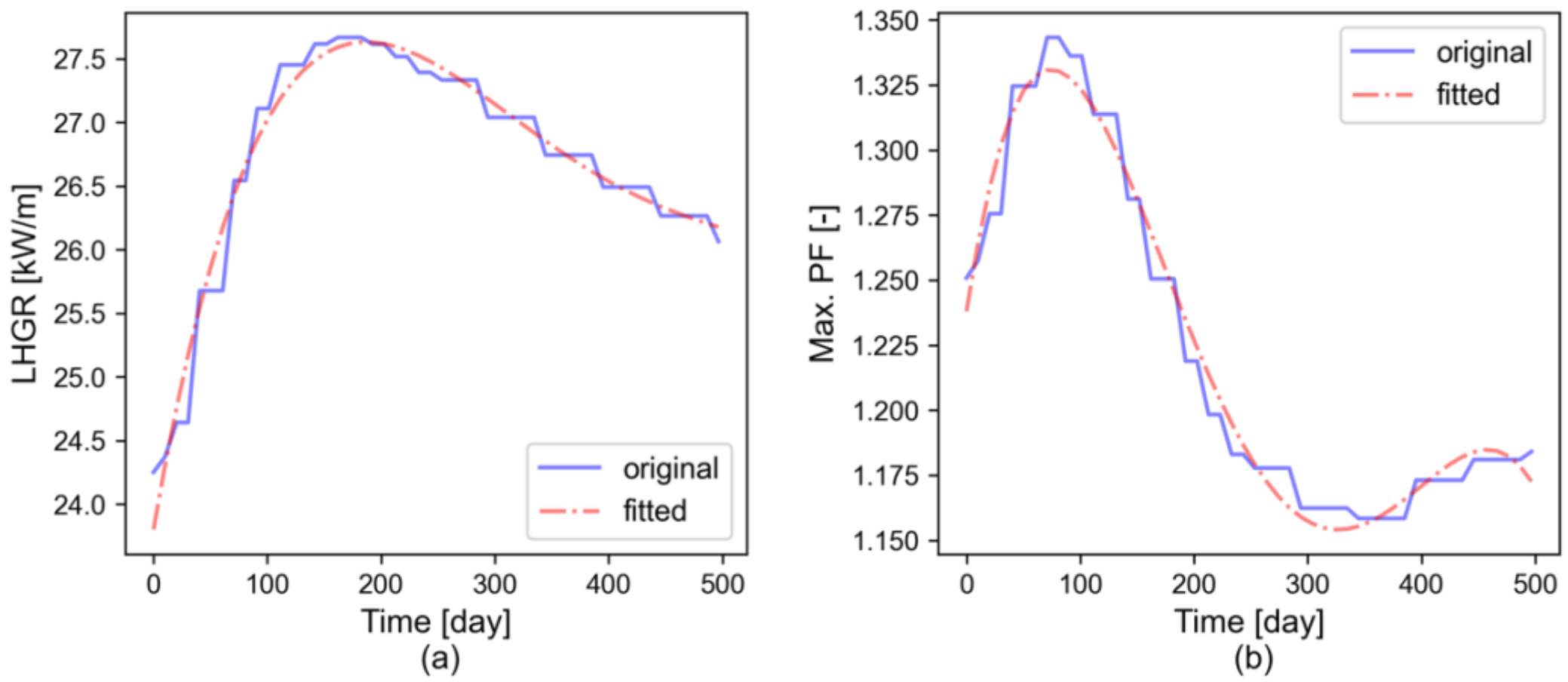

Figure 5. Demonstrative polynomial regression in comparison to the original data in one power cycle of a typical rod, for (a) LHGR time series, and (b) maximum PF time series.

In addition to providing an effective feature set, the rule-based model representation inherently comes with sufficient flexibility hence extensibility. This is critically important as the training data is expected to be designed with good space-filling property within ranges that are physically possible. While the available data at hand is usually limited, the polynomial representation allows creation of synthetic dataset into unseen but physically relevant regions with assistance of computational tools.

## 5.3 Experimental design: extended space-filling design

In the context of surrogate modeling, data collection is typically achieved through computer experiments (or code runs, sampling). Improving the space-filling property not only promotes computational efficiency in sampling, but also ensures the appropriateness of training data such that extrapolation is avoided during generalization. The procedure to come up with an optimal design is often referred to as experimental design, or design of experiments (DOE). In this work, the synthetic training data should sufficiently cover the extended feature space that corresponds to all possible power histories with physical desirability, to make sure that the surrogate is extensible to unseen core patterns. The experimental design is carried out by an extended space-filling design with assistance of clustering and Latin Hypercube sampling (LHS).

For one specific fuel loading pattern, samples fall into certain ranges in the transformed feature space. Such ranges can be very dissimilar for different fuel loading patterns. This work takes three representative fuel loading patterns, and the feasible physical bounds (to the best of existing of knowledge) are defined as the union of the outermost boundaries from the three available cores combined. No effort is made to retrieve the exact pairwise correlation between raw feature for two reasons. First, it becomes a challenging task as dimensionality increases. Second, it is unnecessary to retain all samples within the exact bounds. On the contrary, it is more beneficial to extend beyond the boundaries to accommodate for unobserved but physically desirable samples in the unseen cores patterns. The experimental design for such purpose is carried out as shown in **Algorithm 1**. Details of the experimental design is illustrated and visualized step by step in **Appendix B**.

**Algorithm 1**: Extended space-filling design

1: Combine all available data in the feature space for several representative cores.
2: Divide feature space into multiple subsets (clusters) $\Xi_i, i \in \{1,2,\dots,m\}$, such that the combined samples are relatively uniformly distributed within each region.
3: **For** each subdomain $\Xi_i$:
 Fit marginal probability distributions to each feature variable.
 Perform maximin Latin Hypercube sampling (LHS) according to the fitted marginal distributions. Denote the LHS sample set as $\Phi_i$.
4 Take union of the LHS sample sets $\Phi = \bigcup_{i=1}^{m} \Phi_i$
5 Discard physically undesirable samples from $\Phi$

LHS samples drawn in this way neglect the dependency between feature variables, whereby advantageously extend into unexplored regions with nice space-filling property. The outermost bounds are obtained based on available data to the best of existing knowledge, which can be consistently updated upon availability of more core patterns. The physical desirability in this work takes the principle of LHGR levels between 0 and 30 kW/m, and the maximum PF value no greater than 1.6. Power histories are reconstructed from drawn samples in the transformed feature space, and independent fuel performance analyses are ran at the designed locations to form the synthetic ML training set.

## 5.4 ML algorithms

Supervised learning algorithms of different categories have been selected and compared to capture the fuel performance QoIs. Predictive performance of representative algorithms from linear regression, kernel method, neural network, and ensemble methods are investigated.

**Partial least squares regression (PLS)**
PLS (a.k.a. projection to latent structures) is an efficient regression method built upon covariance. In PLS, predictors are reduced to a smaller set of latent factors (orthogonal) that maximize the explained variance in dependent variables. Least squares regression is then performed between the latent factors and response variables. PLS can be especially useful when the predictors are correlated or when the number of predictors is similar to/higher than observations [14].

**Gaussian process regression (GP)**
GP is introduced into this work as a representative of the kernel method. GP is a nonparametric and probabilistic model that calculates the probability of distribution over all permissible functions that fit the data. As a Bayesian method, a Gaussian process prior needs to be specified in the function space. The form of mean function and covariance function (kernel) in the Gaussian process prior is tuned based on the training data. The posterior distribution is calculated based on the Bayes rule, and the predictive posterior distribution can be obtained on points of interest. The merit of employing GP is that it naturally comes along with uncertainty measurement in the predictive posterior distribution, which benefits the uncertainty quantification procedure in many engineering applications [15]. Nonetheless, the computational cost of training a GP explodes with increased input dimensionality or training size.

**Support-vector regression (SVR)**
SVR is an alternative kernel method that extends from support-vector machines to solve regression problems. The main idea is to first transform inputs to higher dimensional feature space via kernels so that the data becomes separable. A simple linear mapping can then be implemented in the transformed space for regression. Various kernels can be used, among which the radial basis function (RBF) kernel and polynomial kernel are the most powerful. Key advantages of SVR are its effectiveness especially in high dimensional spaces as well as the universal approximation capability [16]. SVRs are more efficient than GPs in general as only a subset of training data is used for support vectors. In the current work, SVR with RBF kernel is employed as an alternative kernel method to GP.

**Neural network (NN)**
As a universal approximator for arbitrary nonlinearity, feed-forward neural network (a.k.a. multilayer perceptron) is a promising ML method. NNs contains layers of interconnected nodes (neurons). Each neuron feeds the weighted sum of all inputs from previous layer to a nonlinear activation function. Hidden layers (intermediate layers between the input and output layer) extrapolate the salient features in the input data that embed predictive power, similar to the effect of feature engineering. Weights and biases are fine-tuned through iterative backpropagation to achieve minimal prediction error on the training data. Under most circumstances, NNs with one or two hidden layers provide sufficient predictive power of input-output nonlinearities. Two NN structures are investigated thereby in this work: two-layered NN (two hidden layer) and three-layered NN (three hidden layers).

**Random forest (RF)**

RF is a popular tree-based ensemble learning algorithm for its simplicity, efficiency and robustness. Individual decision trees are trained in parallel on different subsets of training data based on unique combinations of features. Result from these decision trees are aggregated together for prediction with reduced variance, so RFs usually outperform decision trees and exhibit good generalization. RFs are easy to tune, robust to noises and training sample selection, and unlike other distance-based ML algorithms, do not require feature scaling. In context of fuel performance modeling, many physical processes are triggered above certain threshold, so the tree-based algorithms are expected to more accurately capture such discontinuities.

**Gradient boosted tree (XGB)**
Gradient boosted trees is another tree-based ensemble method that is gaining wide use. The key idea of "boosting" is to iteratively combine "weak learners" into one single "strong learner". Functional gradient descent algorithm is used to optimize the cost function over the function space. Similar to RFs, boosted trees use an ensemble of decision trees for prediction, and feature scaling is not necessary. Unlike RF which combines a forest of random trees in parallel, decision trees in boosted trees are trained successively on the error of previous trees. A series of shallow trees are ensembled in a nonrandom way, and the prediction error becomes smaller as more trees are added. Making predictions with boosted trees is fast and not memory-consuming once the model is built. Python package XGBoost is employed to implement the gradient boosted tree, and it is referred to as XGB (extreme gradient boosting) in this work [17].

## 6. Results and discussion

ML algorithms are implemented from Keras [18], Scikit-learn [19], and XGBoost [17] in Python 3.6. The dataset is partitioned into a training set (on which the ML model is trained) and a test set (on which the predictive performance of the model is evaluated). $k$-fold cross-validation is used for model selection. The training set is randomly shuffled prior to cross-validation, and is averagely split into $k$ folds. Each one of the $k$ folds is held out once to evaluate the performance of model trained on the rest $(k-1)$ folds. Mean and standard deviation of the model evaluation scores are stored as the assessment of model effectiveness. The test set is a holdout data set that has never been used for training, and the predictive performance of the trained model is evaluated on the test set. This section presents the predictive performance of ML surrogates in comparison to LUT for the two tasks, with $k = 5$.

### 6.1 Predictive performance of ML surrogates

In this work, a total number of 9128 samples are drawn from the extended space-filling design as training set. Similarly, five-fold cross-validation is performed to select the best model structure for each QoI. The predictive performance is tested on a total number of 16,632 realistic fuel pins from three distinguished cores.

The training set in this case is purely synthetic from experimental design on extended data range to accommodate unobserved cores. LHGR time series are reconstructed from polynomial coefficients in the feature space, as shown in Figure 6. Although rods in real reactors may never experience some of the power histories in Figure 6, adding training samples at such locations help prevent extrapolation of the ML algorithms when predicting on unexplored core patterns. Core-average PFs are used for PF profiles to preserve the location of maximum peaking factors. FRAPCON inputs are generated and ran based on the reconstructed operating conditions for training data collection.

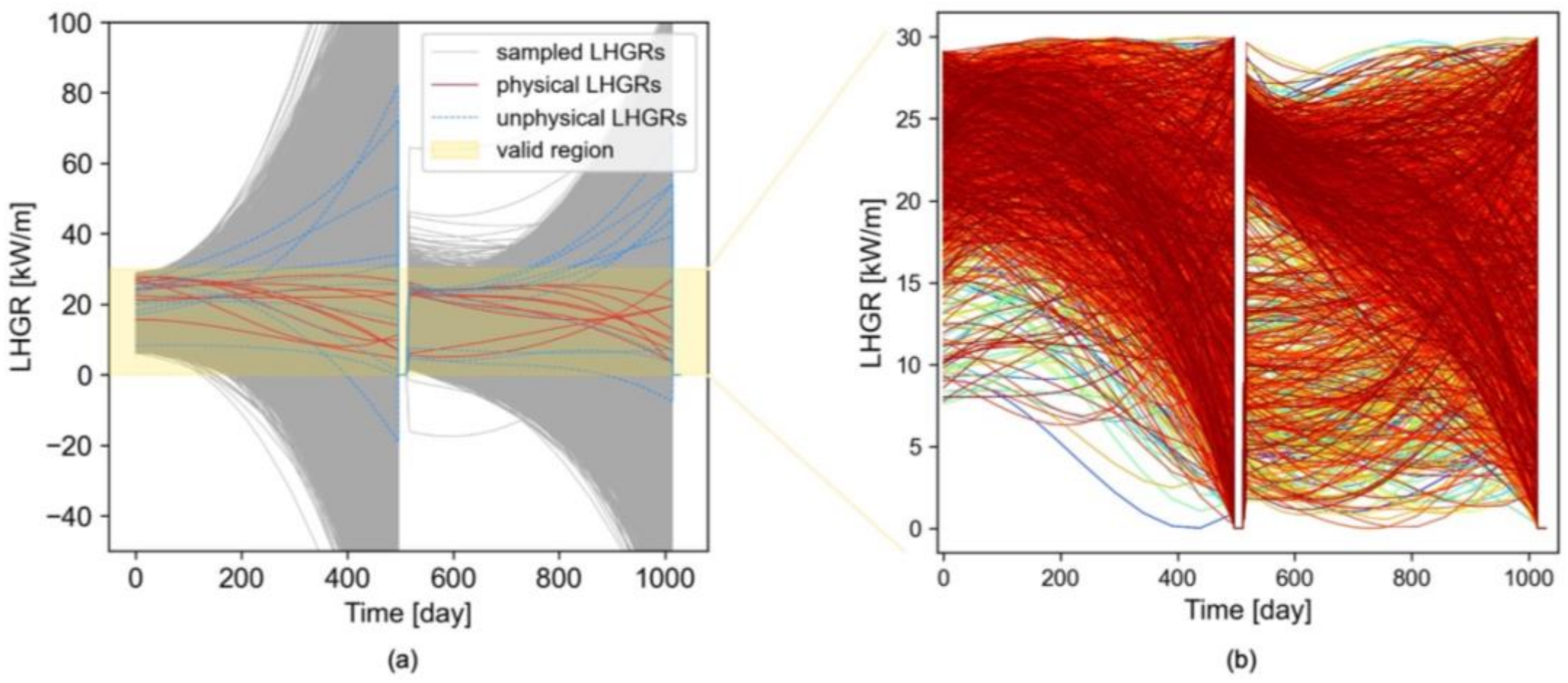

Figure 6. Reconstructed power histories in the extended space-filling design. In (a), gray lines represent all sampled LHGRs. Red/blue lines are a few representatives of the physical/unphysical LHGR samples, where physical LHGR time series are those falling into the “valid region” (between 0 and 30 kW/m). All physical LHGR time series that are used as training set in this work in shown in (b).

Despite superior flexibility, the spatial-temporal operating condition reconstructed in this way inevitably loses details. Figure 7 shows the reconstruction errors for the eight QoIs on the three available cores with a total number of 16,632 fuel pins. Such reconstruction errors result from a compromise between dimensionality of the feature space and reconstruction precision on the spatial-temporal dependent PF profiles. Presently the PF profile at each timestep is represented with the maximum PF value. More raw features can be employed for more accurate reconstruction of the PF profiles, while the improvement in reconstruction accuracy does not compensate for the loss of accuracy in ML training due to increased input dimensionality. Such reconstruction error is a necessary sacrifice to make and serves as a primary source of prediction error in the ML surrogates, as confirmed in Figure C.1 in **Appendix C** which compares the surrogate predictive error versus the reconstruction error with the best selected surrogate.

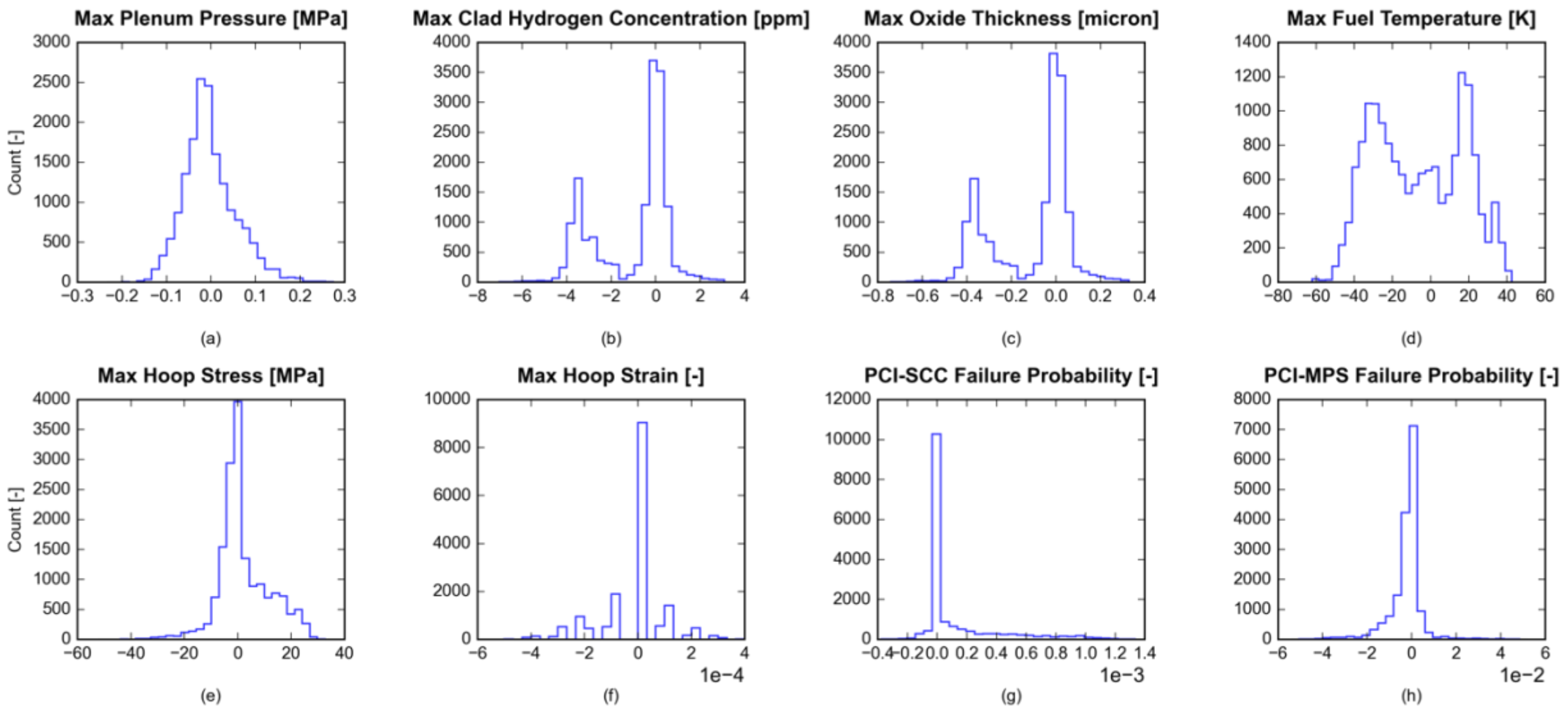

Figure 7. Reconstruction errors for the eight fuel performance QoIs .

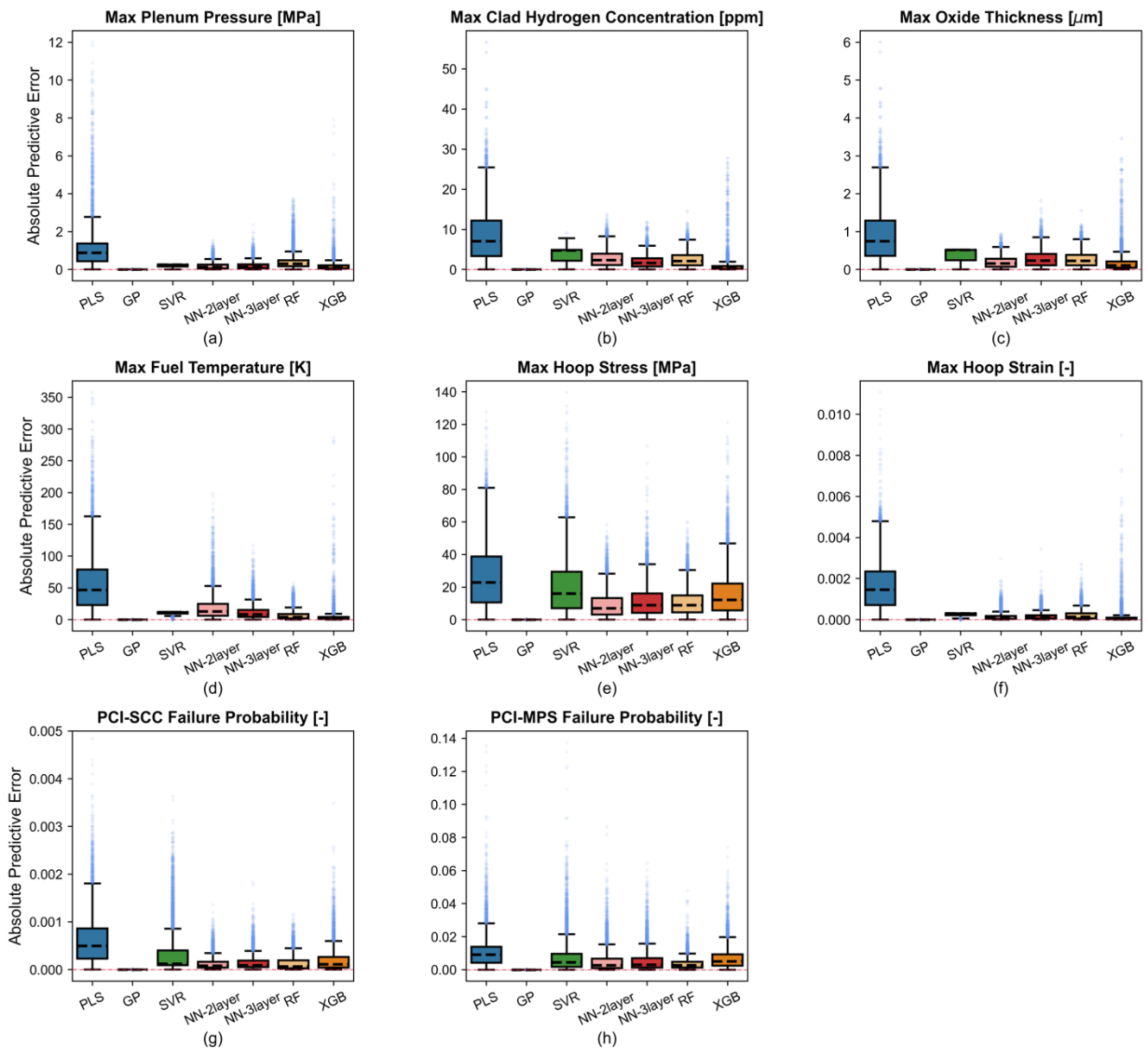

Figure 8. Boxplots of absolute predictive error with LUT and ML algorithms on the synthetic training set for the eight fuel performance QoIs.

Figure 8 shows the absolute error on the training set with different ML algorithms for each fuel performance QoI. The predictive capability of aforementioned ML algorithms on the three unseen cores is compared against LUT for each fuel performance QoI, as shown in the boxplots of absolute predictive errors in Figure 9. Figure 10 (a) to (f) plots the cumulative data fraction versus relative absolute error, with the relative root-mean-square error (rRMSE[1]) shown in the side table. The rod failure probabilities in Figure 10 (g) and (h) are by definition small fractional numbers, under which case the absolute error and correspondingly mean absolute percentage error (MAPE) are used as more sensible error metrics. All subplots in Figure 10 are horizontally zoomed in for better visibility.

A more detailed comparison of the surrogate performance is summarized in Table C.1 in **Appendix C**, including the predictivity coefficient $R^2$, RMSE, MAE, maximum and minimum prediction error on the test set. $R^2$ compares the variation in the predictive error to that of the observations. RMSE and MAE are both

[1] Relative root-mean-square error (rRMSE) is defined as

$$\text{rRMSE} = \sqrt{\frac{1}{N}\sum_{i=1}^{N}\left(\frac{\hat{y}_i - y_i}{y_i}\right)^2}$$

calculated in this work, with MAE intended for average predictive error, and RMSE accounting for variance of the predictive error[2]. RMSE is more sensitive to outliers than MAE due to quadratic scoring, and the difference in the model rankings between the two metrics indicates presence of outliers. In addition, the maximum and minimum predictive errors are also extracted for outlier identification.

The best $R^2$ for the maximum plenum pressure, maximum hydrogen concentration, maximum oxide thickness, and maximum fuel temperature are all able to reach 0.95, indicating excellent predictivity. The maximum hoop stress is subject to greater prediction error due to the inherent uncertainty of its occurrence as well as relatively larger reconstruction error, while the best $R^2$ is still above 0.92. For some rods that go through high LHGR, the maximum cladding hoop stress occurs in the form of a spike at the beginning of the second power cycle due to PCI, leading to large tensile (positive) hoop stress. PCI may never take place in some other rods, so the maximum hoop stress remains compressive (negative) throughout the life. It is possible that other rods undergo relatively moderate PCI, such that the maximum hoop stress happens at some time during a power cycle. Therefore, the maximum hoop stress covers a wide range, from negative values (typically around -70 MPa) to positive numbers (up to 190 MPa). The relative errors are magnified for small absolute values (close to zero), which explains the large rRMSE values (over 100%) as shown in Figure 10(e). The RMSE and MAE are 15.3 MPa and 12.4 MPa for the best surrogate. It is also worth mentioning that the accurate prediction of high tensile hoop stress matters more than that of compressive hoop stress, as peak tensile hoop stress directly correlates to PCI failure risk. The maximum hoop strain is subject to similar uncertainties and reconstruction errors, and the best $R^2$ is approximately 0.89.

The last two QoIs (failure probabilities due to PCI-SCC or PCI-MPS) are generally harder to predict due to more complicated physics, as damage accumulation is triggered by the maximum hoop stress above certain temperature and burnup-dependent threshold, and the prediction of maximum hoop stress itself is subject to relatively large error. Regarding MAE which is a more physical error metric for the last two QoIs, the averaged prediction error is only 0.022 % and 0.67% for PCI-SCC and PCI-MPS failure probability. The predictivity for these two QoIs are therefore taken as acceptable.

LUTs perform as good as NNs for the maximum plenum pressure, and outperforms all ML models for the maximum fuel temperature. For the remaining QoIs, ML models achieves better generalization than LUTs. NNs excel over other ML algorithms due to better robustness to reconstruction error for all QoIs.

where $y_i$ and $\hat{y}_i$ are the observation and surrogate prediction.
[2] RMSE is always greater equal than MAE by definition.

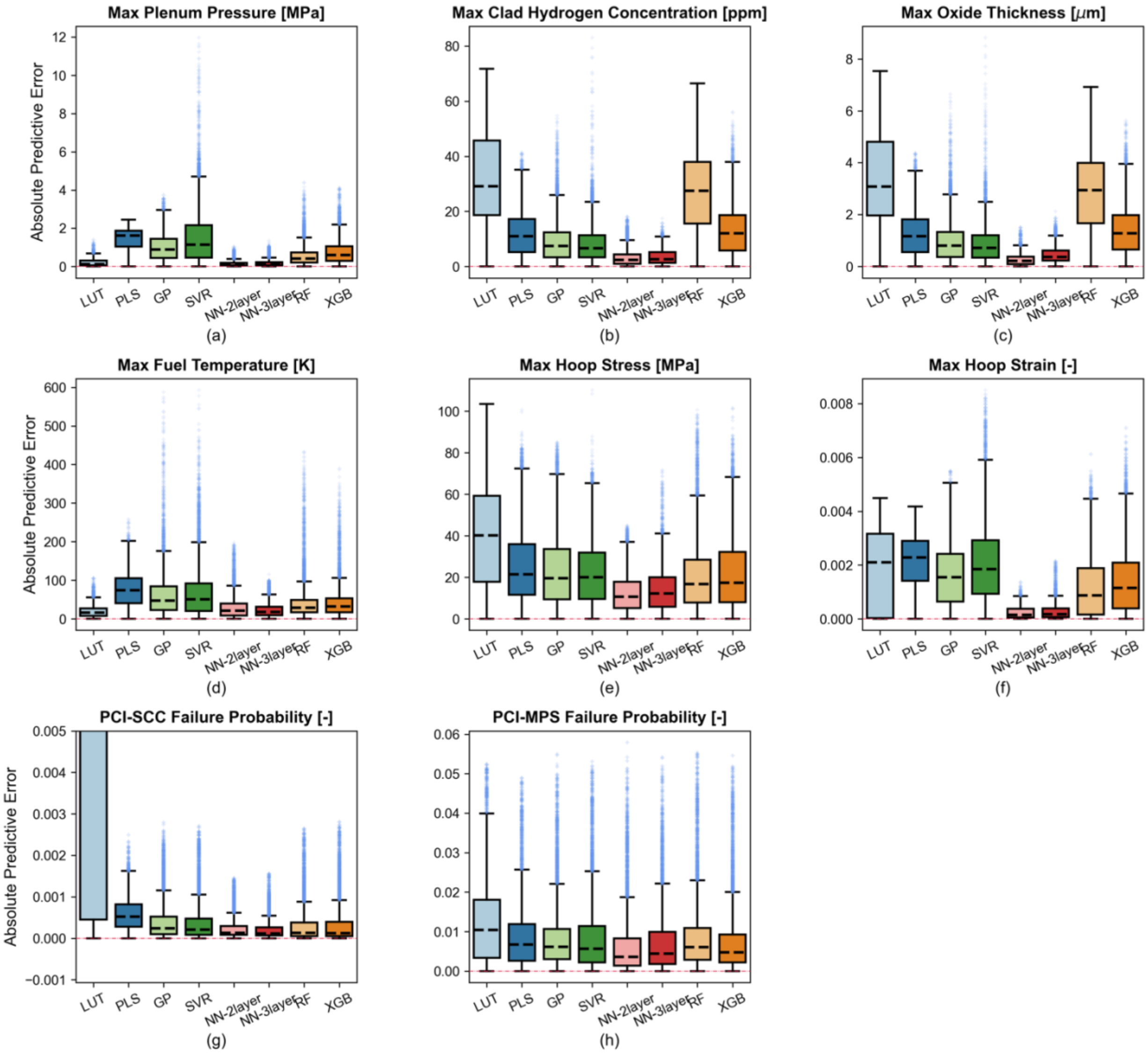

Figure 9. Boxplots of absolute predictive error with LUT and ML algorithms on the three unseen cores for the eight fuel performance QoIs.

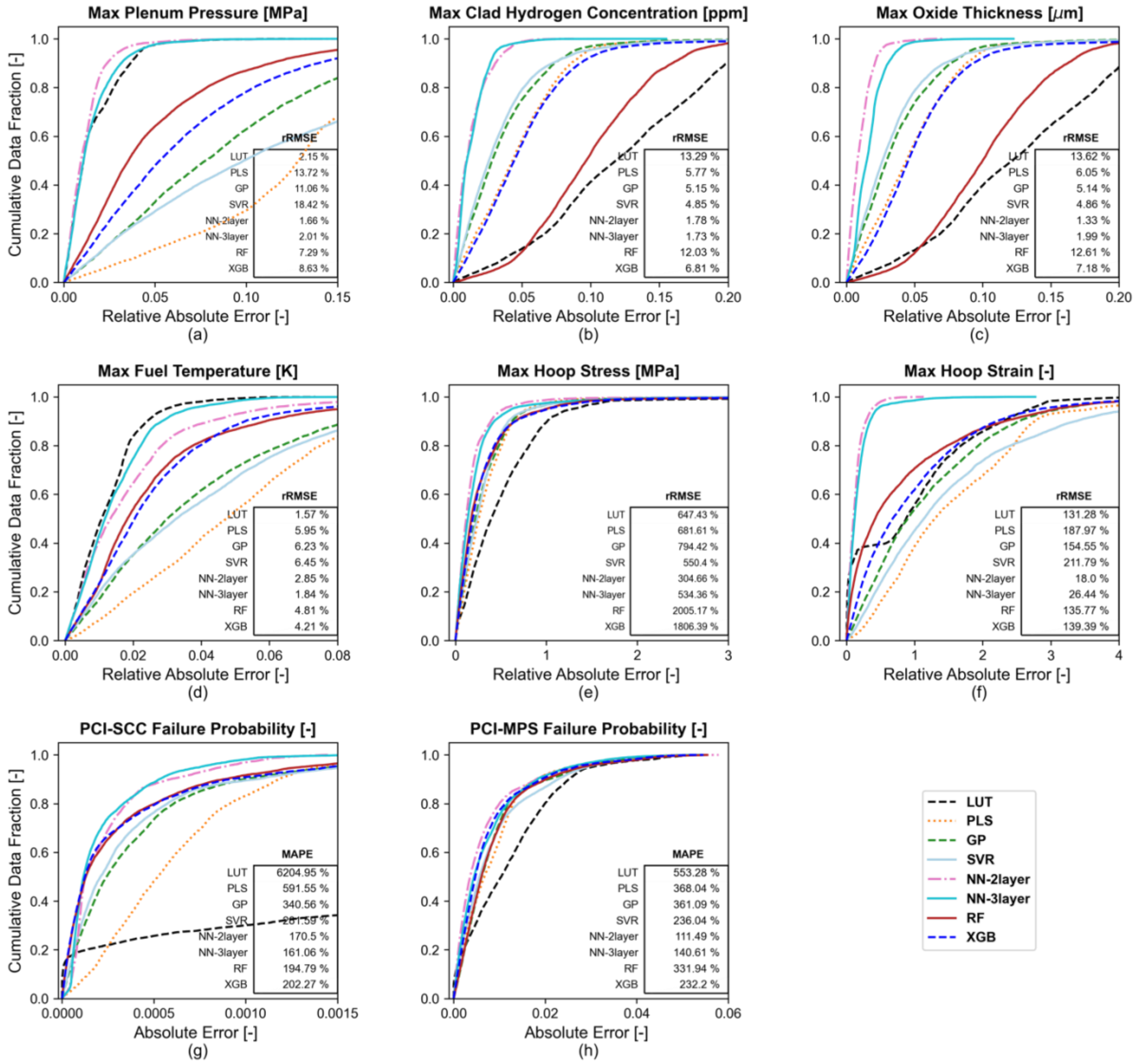

Figure 10. Cumulative data fraction of (relative) absolute errors for LUT and ML algorithms for the eight fuel performance QoIs.

## 6.2 Runtime analysis

With the goal of speeding up the full-core fuel performance modeling, the computational complexity (especially time complexity) during surrogate construction and prediction need to be quantified. Let $n$ be the total number of training samples and $d$ represent the input dimension (in this work $d \ll n$). The asymptotic worst-case/average-case time complexity[3] during training and prediction for different algorithms are summarized in Table 6.1[4]. The training complexity describes the worst-case/average-case runtime to train an algorithm on the training set containing $n$ samples. The effort devoted to data preparation and model selection is not included in the training complexity. The prediction complexity is the runtime consumed for making prediction on a single test point. LUT is a constructed as a 2D matrix for repeated use in this work, hence the training complexity is not applicable to LUT. For NN, the time complexity is structure-dependent so only the prediction complexity is estimated.

[3] Big-O or $\mathcal{O}$ is the asymptotic notation for the worst-case runtime. Big Theta or $\Theta$ denotes the asymptotically tight bound on the runtime.

[4] Note that analytic form of the time complexity depends on specific implementation of the algorithm. The analyses shown in Table 6.1 are intended for a general comparison.

Table 6.1. Time complexity for the surrogate

| Surrogate | Training Complexity | Prediction Complexity | Nomenclature |
|---|---|---|---|
| **LUT** | - | $\mathcal{O}(1)$ | |
| **PLS** | $\mathcal{O}(nd^2)$ | $\mathcal{O}(d)$ | |
| **SVR** | between $\mathcal{O}(n^2p)$ and $\mathcal{O}(n^3p)$ | $\mathcal{O}(n_{sv}p)$ | $p$: number of features<br>$n_{sv}$: number of support vectors |
| **GP** [20] | $\mathcal{O}(n^3)$ | $\mathcal{O}(n)$ | |
| **NN** | - | $\Theta(dn_{l_1} + n_{l_1}n_{l_2} + \cdots)$ | $n_{l_i}$: number of neurons in $i$-th hidden layer |
| **RF** [21] | $\Theta(mkn\log(n))$ | $\Theta(m\log(n))$ | $m$: number of random trees<br>$k$: number of features considered for splitting |
| **XGB**[17] | $\mathcal{O}(mhn\log(n))$ | $\mathcal{O}(mh)$ | $m$: total number of trees<br>$h$: maximum depth of the tree |

Despite that the runtime depends on the training size and model structures, the training/prediction efficiency can be compared in the context of this application, as shown in Table 6.2. Note that the "training time" in Table 6.2 stands for the one-time training time, while in reality such training procedure needs to be repeated a lot of times during hyperparameter tuning. It is also worth mentioning that the runtime of each algorithm listed in Table 6.2 also depends on the implementation efficiency of the python library used. Although LUT does not run the fastest, it has extensive applicability without the need to be reconstructed for new operating conditions. PLS runs the fastest as a projection plus linear regression method. Tree-based algorithms are also fast in prediction, with XGB being more efficient than RF during both training and prediction. NNs are relative slow for both training and prediction, as numerous hyperparameters embedded in NNs increase the computational cost during both model selection and prediction. GP is the most computationally expensive among all algorithms. In comparison to FRAPCON which takes approximately 5.8 seconds for two-cycle rods, the combined surrogate that contains the best surrogate for each QoI (Table 6.3) achieves around 4000 times acceleration as listed in the last column of Table 6.2. This provides great potential to assist the core design optimization procedure where large number of patterns are need to be investigated.

Table 6.2. Average training time and runtime for different algorithms

| Surrogate | Training time [s]<br>(training size 9128) | Runtime [μs]<br>(on a single rod) | Runtime Acceleration<br>(FRAPCON runtime / surrogate runtime) |
|---|---|---|---|
| **LUT** | - | 186.910 | 3.103E+04 |
| **PLS** | 0.0288 | 0.186 | 3.103E+07 |
| **SVR** | 224.867 | 233.376 | 2.485E+04 |
| **GP** | 930.980 | 3240.176 | 1.790E+03 |
| **NN-2layer** | 0.900 | 29.863 | 1.942E+05 |
| **NN-3layer** | 1.015 | 34.294 | 1.691E+05 |
| **RF** | 67.915 | 174.032 | 3.333E+04 |
| **XGB** | 36.077 | 22.117 | 2.622E+05 |
| **Combined surrogate*** | 5.515 | 557.428 | 1.040E+04 |

*The combined surrogate composes of the best surrogates as shown in Table 6.3 for prediction of all QoIs at the same time.

## 6.3 Selected surrogate for each fuel performance QoI

Selection of the most desirable surrogate requires comprehensive consideration of prediction accuracy and efficiency. The judge on prediction accuracy prefers not only high predictivity coefficient $R^2$, but also low RMSE, MAE and fewer outliers. When the prediction precisions come out comparable for two algorithms, it is more ideal to select the one with simpler structure or the one that runs faster. Particularly, it is always rewarding to employ LUT as surrogate when it is capable to provide sufficient predictive power, as LUT is of more general applicability compared to other ML algorithms. Out of these considerations, the recommended surrogates are shown in Table 6.3. LUT and NNs are generally more preferable. Predictive performance versus true observations with the selected best surrogates for the 2-cycle rods are shown in Figure 11.

Table 6.3. Best surrogate for fuel performance QoIs.

| QoIs | 2-cycle rods Best Surrogate | 2-cycle rods $R^2$ (*MAE) | 3-cycle rods Best Surrogate | 3-cycle rods $R^2$ (*MAE) |
|---|---|---|---|---|
| **Max Plenum Pressure [MPa]** | LUT | 0.972 | LUT | 0.960 |
| **Max Hydrogen Concentration [ppm]** | NN-2layer | 0.995 | NN-2layer | 0.985 |
| **Max Hoop Stress [MPa]** | NN-2layer | 0.922 | NN-3layer | 0.904 |
| **Max Hoop Strain [-]** | NN-2layer | 0.899 | NN-2layer | 0.789 |
| **Max Oxide Thickness [µm]** | NN-2layer | 0.997 | NN-2layer | 0.98 |
| **Max Fuel Temperature [K]** | LUT | 0.963 | LUT | 0.912 |
| **PCI-SCC Failure Probability [-]** | NN-3layer | *0.022% | NN-3layer | 0.0078% |
| **PCI-MPS Failure Probability [-]** | NN-2layer | *0.67% | NN-3layer | 0.64% |

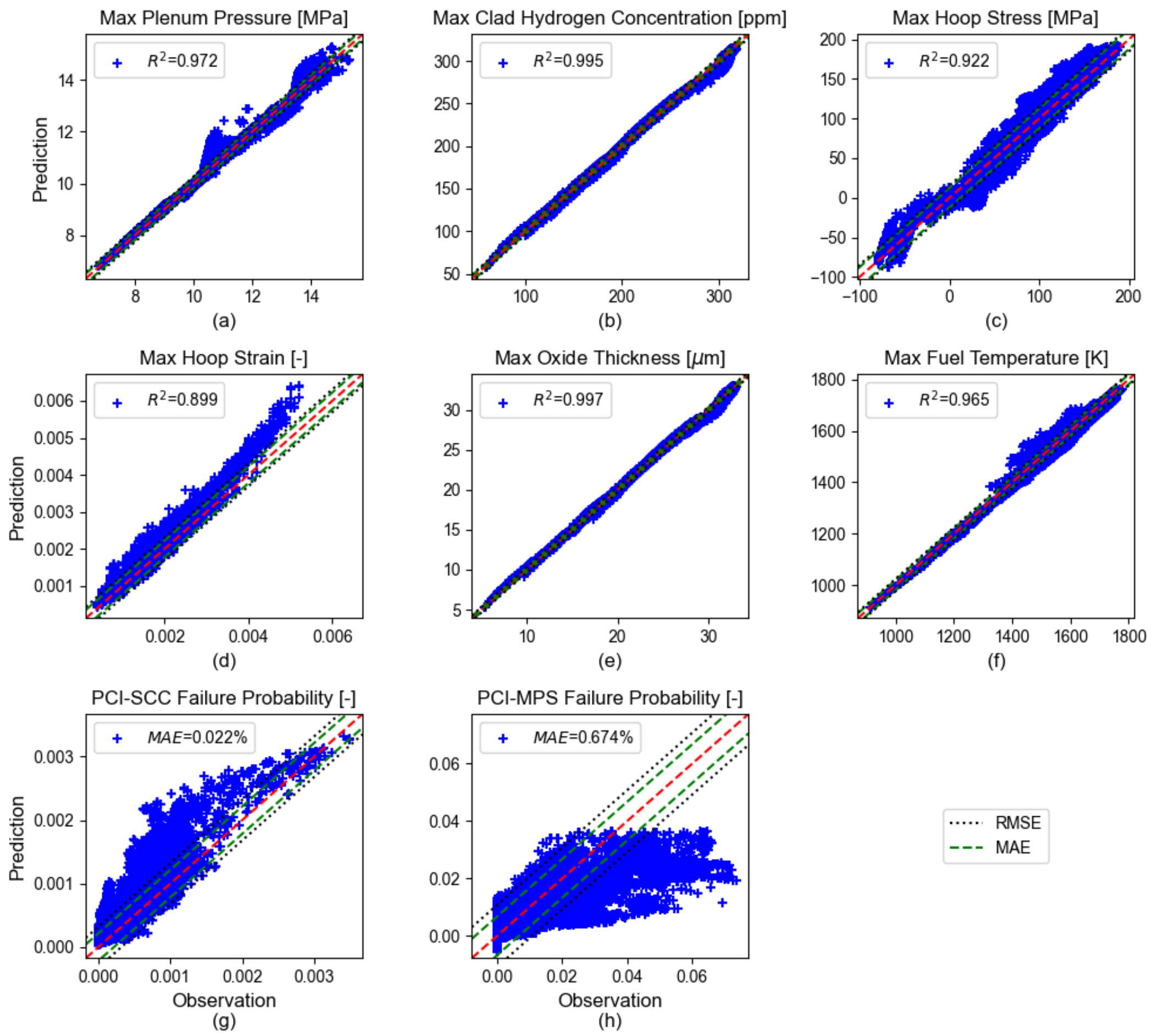

Figure 11. Predictions versus observations with the best surrogate for each fuel performance QoI.

## 7. Conclusions

This work provides a first study on construction of fast-running surrogates for essential fuel performance QoIs to accelerate the full-core fuel performance analysis. Constructed surrogates efficiently reduce the runtime by at least three orders of magnitude compared to fuel performance analysis code FRAPCON, allowing for direct coupling of the fuel performance responses into the core design optimization. For certain fuel performance QoIs that embed complicated physics, ML models compensate for the inadequacy of conventional LUTs. Different techniques for feature engineering and experimental design are introduced and used, depending on the required applicability of the surrogate. The best surrogate can be thereby selected based on overall consideration of both prediction accuracy and time complexity. A comprehensive study is provided to demonstrate the application and applicability of LUTs versus different ML models for each fuel performance QoI, and the following conclusions can be drawn from this work:

- As an feature extraction technique, rule-based model offers enough flexibility hence extensibility to unseen data range.
- In order to extend the applicability of ML algorithms to unobserved core patterns, an extended space-filling design is performed on the enlarged domain to make sure that the training data sufficiently covers the data range with physical desirability.
- With operating power as the only variating condition, the maximum fuel temperature and maximum oxide thickness is highly linear with respect to burnup and power level. Therefore LUTs show superiority in both accuracy and simplicity.
- For other fuel performance QoIs that involve more complicated physics, ML-assisted surrogates outperform LUTs in prediction accuracy.
- NNs are powerful to capture the nonlinear input/output relation and are more robust to noise (reconstruction error) compared to other ML algorithms, therefore favored in the current analysis.
- Regarding the time complexity of the surrogate, the cost of training a ML surrogate reduces in the order of kernel methods (GP and SVR), tree-based algorithms (RF and XGB), neural networks and linear regression (PLS) in a general way. The prediction efficiency improves from GP, SVR, LUT, RF, NN and XGB, to PLS. Linear regression (PLS) is not suitable for most of the multiphysics problems, but provides a good benchmark.

Beyond the study conducted in this paper, potential efforts can be made in the future work:

- Updating physical range for the extended space-filling design, upon availability of more fuel loading patterns.
- Increasing the model complexity by taking into consideration more physical parameters, e.g. number of batches, power cycle length, load following etc. Inclusion of such scalar variables likely increase the input dimension without adding extra burden to the overall workflow.
- Generalizing the full-core surrogate to high-fidelity fuel performance code like BISON for more accurate full-core monitoring.
- Integrating the fast running surrogate with core reload optimization to achieve tighter coupling of fuel performance feedbacks to other physics.

## Appendix A. Calculation of rod failure risk

Under the low temperature steady-state operating condition, a cumulative damage model [22] [10] can be used to estimate the cladding damage induced by stress corrosion cracking during PCI. The amount of damage (cumulative damage index, CDI) at time $t_n$ is defined by

$$\mathrm{CDI} = \int_0^{t_n} \frac{dt}{t_f(\sigma, Bu, T)}$$

where $t_f$ is the time-to-failure (in seconds) as a function of the clad hoop stress $\sigma$ (MPa), burnup $Bu$ (MWd/MTU) and temperature $T$ (K). The time-to-failure $t_f$ is calculated via

$$t_f = \bar{t}\exp[(1.015\sigma_y + 1.74\sigma_{ref} - 2.755\sigma) \times 10^{-2}]$$

where

$$\bar{t} = 5 \times 10^5 (1.13 \times 10^{-4} Bu - 0.13)^{-0.75} \exp(-30(1 - 611/T))$$

$\sigma_{ref}$ is a threshold stress (MPa) in form:

$$\sigma_{ref} = \begin{cases} 336.476(Bu - 5000)^{-0.07262} & Zr2 \\ 310.275(Bu - 5000)^{-0.04400} & Zr4 \end{cases}$$

Cladding yield stress $\sigma_y$ (MPa) as a function of cladding temperature is interpolated according to [23]. The CDI model is only activated when hoop stress exceeds the threshold stress, and burnup greater than 5000 MWd/MTU. The damage accumulation time is limited to 1000 seconds upon reaching the peak hoop stress based on in-reactor PCI observations, as suggested in [11].

To account for the local stresses and keep consistent with the developed failure metrics, the 2D radial-axial (R-Z) hoop stress is mapped to radial-azimuthal (R-$\theta$) through stress concentration factors (CFs) [11], estimated as:

$$CF_{\mathrm{PCI-SCC}} = -0.0042\sigma_\theta + 2.3773$$
$$CF_{\mathrm{PCI-MPS}} = -0.0115\sigma_\theta + 4.3099$$

The cumulative distribution function for the cumulative damage index is described as below. The rod failure risk can be statistically determined from the cumulative distribution function, and rods with failure probability greater than 0.5 are vulnerable to PCI failures. The procedure to calculate the rod failure risk is summarized in Algorithm A.1.

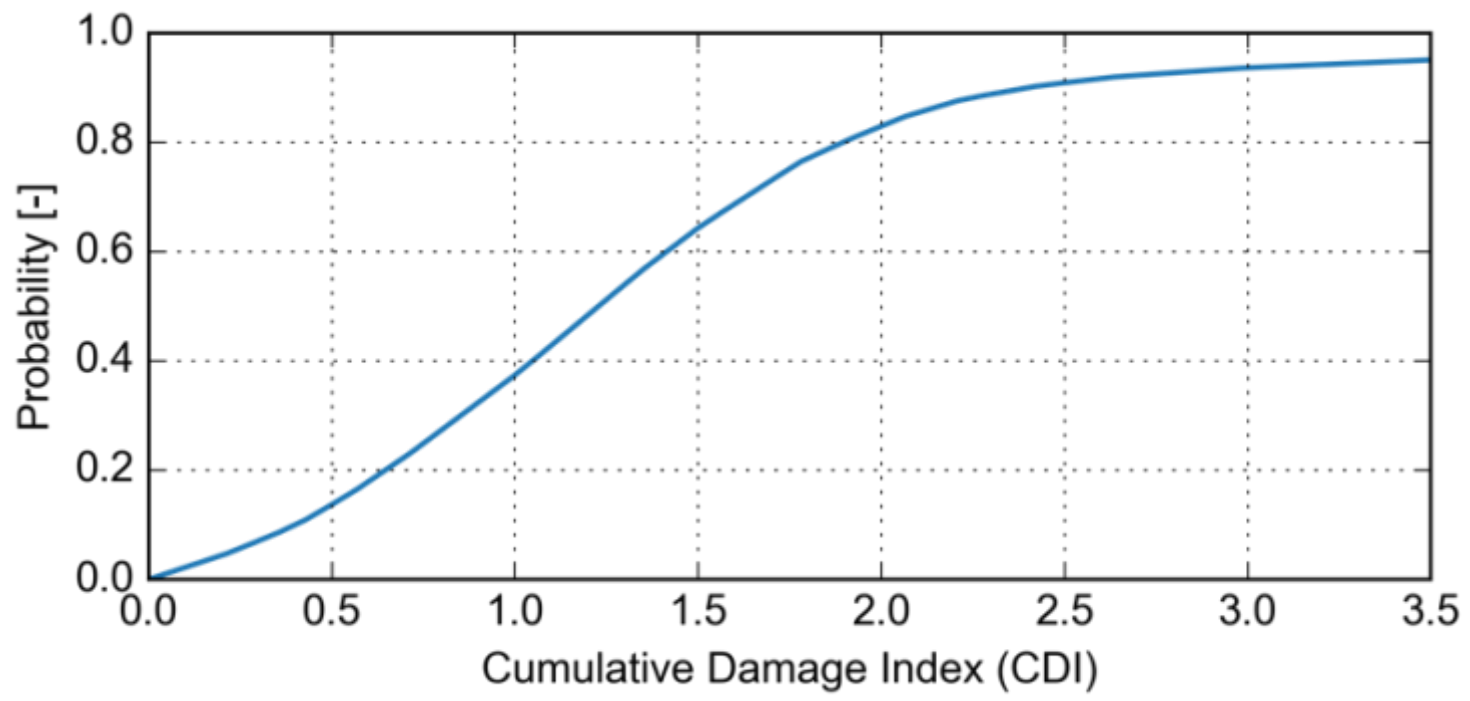

Figure A.1. Cumulative distribution function of CDI.

| **Algorithm A.1**: Fuel rod failure risk calculation | |
|---|---|
| 1: | Perform fuel performance analysis and identify the peak R-Z hoop stress $\sigma_\theta$. |
| 2: | Apply the stress concentration factor for PCI-SCC and PCI-MPS to the R-Z hoop stress $\sigma_\theta$. |
| 3: | Calculate CDI for PCI-SCC and PCI-MPS. |
| 4 | Infer the rod failure risk from the cumulative distribution function of CDI according to Figure A.1. |

## Appendix B. Experimental design with extended space-filling design

The extended space-filling design is used to create the synthetic training set to predict unseen core patterns. Figure B.1 (a) to (d) show the procedure step by step based on three available equilibrium PWR cores. First, the rule-based model (polynomial representation) is used for feature extraction, with the pairwise distribution in the feature space shown in Figure B.1 (a) for the first seven features for better visibility. The feature space is then divided into local regions by clustering (Figure B.1 (b)). Figure B.1 (c) plots the LHS samples generated according to the fitted marginal distributions, with the original core data visualized on top of the LHS samples. The LHS samples retain the same data range as the original core data with nice space-filling property in the probability space. Finally, LHS samples that correspond to unphysical operating conditions are discarded, and Figure B.1 (d) presents the physical samples versus the full LHS design. In the current study, 16,632 two-cycle rods are available from the three PWR cores. The feature space is divided into three clusters, and 8,000 samples are generated within each region. Upon screening out the unphysical samples, a total number of 9,128 samples are left and used as training set for the ML algorithms.

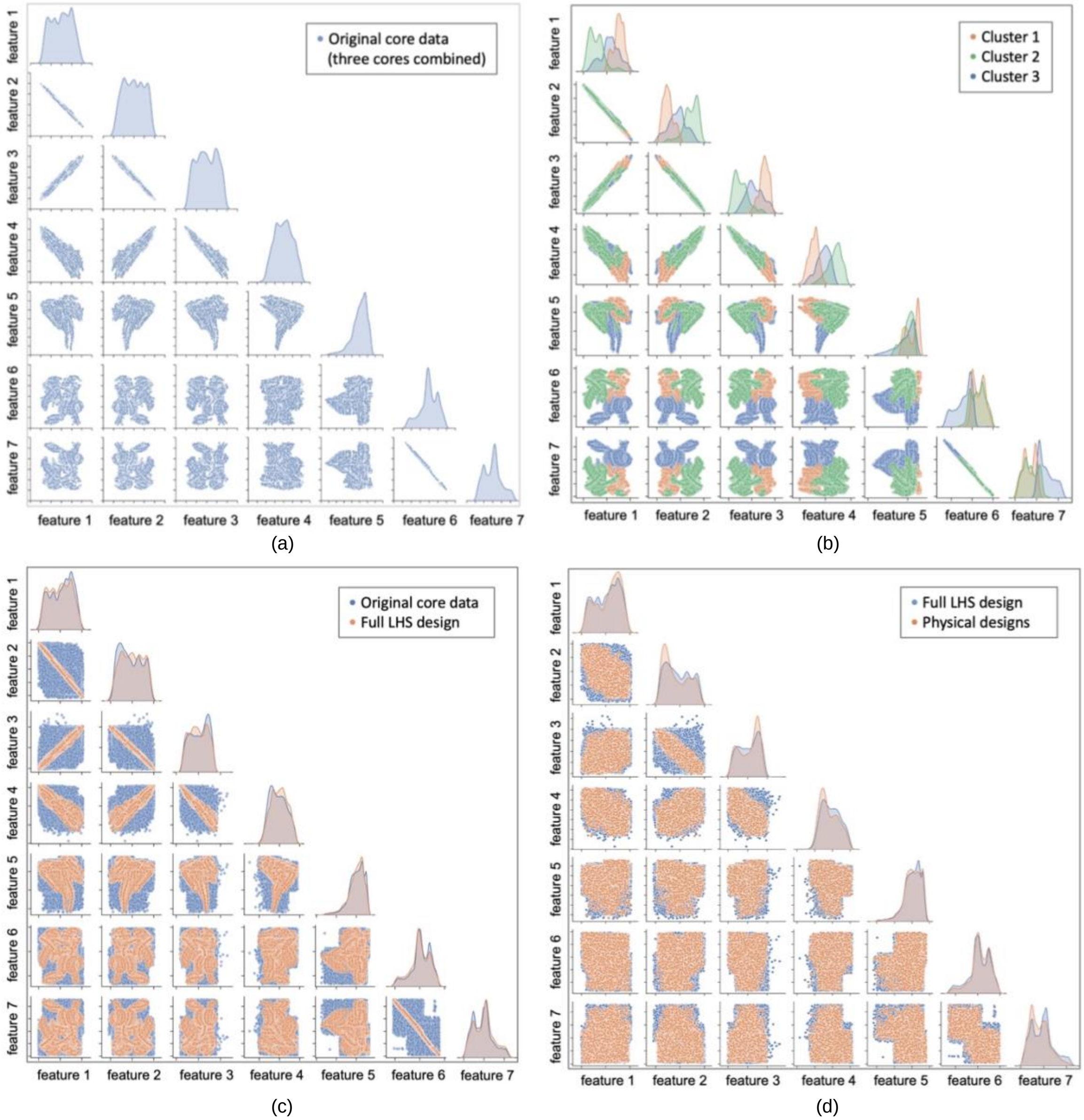

Figure B.1. Extended space-filling design on the rule-based model feature space. (a). pairwise distribution of the features for three available cores. (b). clustering of the features. (c). LHS samples generated according to the marginal probability distributions within each cluster. (d). physical samples after ruling out the unphysical samples. Only the first seven features are shown for simplicity.

## Appendix C. Predictive performance of LUT and ML surrogates in detail

Detailed comparison of the surrogate performance is summarized for 2-cycle rods in Table C.1 and for 3-cycle rods in Table C.2, including the predictivity coefficient $R^2$, RMSE, MAE, maximum and minimum prediction error on the test set (three unseen cores). Figure C.1
Figure C.2 and Figure C.3 show the surrogate performance on the test set for all the 3-cycle rods in the three available cores.

Table C.1. Predictive performance of LUT vs. ML algorithms for 2-cycle rods (green denotes the best desirability, and blue indicates the worst performance)

| | | Max Plenum Pressure [MPa] | Max Hydrogen Concentration [ppm] | Max Hoop Stress [MPa] | Max Hoop Strain [-] | Max Oxide Thickness [μm] | Max Fuel Temperature [K] | PCI-SCC Failure Probability [-] | PCI-MPS Failure Probability [-] |
|---|---|---|---|---|---|---|---|---|---|
| $R^2$ | LUT | 0.972 | 0.738 | -0.044 | -5.353 | 0.742 | 0.963 | -476.882 | -0.385 |
| | PLS | 0.053 | 0.946 | 0.682 | -4.541 | 0.946 | 0.542 | -1.609 | 0.338 |
| | SVR | -0.676 | 0.971 | 0.755 | -5.876 | 0.971 | 0.484 | -0.905 | 0.207 |
| | GP | 0.400 | 0.965 | 0.721 | -3.031 | 0.964 | 0.584 | -0.845 | 0.312 |
| | NN-2layer | 0.986 | 0.995 | 0.922 | 0.899 | 0.997 | 0.907 | 0.374 | 0.462 |
| | NN-3layer | 0.980 | 0.995 | 0.901 | 0.842 | 0.993 | 0.955 | 0.486 | 0.427 |
| | RF | 0.779 | 0.755 | 0.768 | -1.702 | 0.753 | 0.779 | -0.481 | 0.262 |
| | XGB | 0.626 | 0.935 | 0.755 | -1.973 | 0.935 | 0.806 | -0.657 | 0.372 |
| RMSE | LUT | 0.263 | 35.332 | 53.927 | 2.322E-03 | 3.721 | 22.435 | 9.283E-03 | 1.649E-02 |
| | PLS | 1.545 | 14.443 | 30.976 | 2.298E-03 | 1.527 | 86.221 | 7.214E-04 | 1.194E-02 |
| | SVR | 2.055 | 10.487 | 27.168 | 2.560E-03 | 1.115 | 91.580 | 6.164E-04 | 1.307E-02 |
| | GP | 1.229 | 11.646 | 29.013 | 1.960E-03 | 1.250 | 82.187 | 6.065E-04 | 1.217E-02 |
| | NN-2layer | 0.187 | 4.521 | 15.312 | 3.099E-04 | 0.334 | 38.920 | 3.534E-04 | 1.077E-02 |
| | NN-3layer | 0.225 | 4.330 | 17.253 | 3.885E-04 | 0.534 | 27.091 | 3.201E-04 | 1.111E-02 |
| | RF | 0.746 | 30.714 | 26.441 | 1.605E-03 | 3.279 | 59.953 | 5.434E-04 | 1.261E-02 |
| | XGB | 0.971 | 15.872 | 27.194 | 1.683E-03 | 1.684 | 56.135 | 5.749E-04 | 1.163E-02 |
| MAE | LUT | 0.194 | 29.684 | 45.753 | 1.759E-03 | 3.132 | 17.684 | 7.112E-03 | 1.268E-02 |
| | PLS | 1.421 | 11.964 | 25.305 | 2.096E-03 | 1.264 | 75.035 | 5.974E-04 | 8.694E-03 |
| | SVR | 1.518 | 8.123 | 22.181 | 2.084E-03 | 0.860 | 66.285 | 3.862E-04 | 8.929E-03 |
| | GP | 1.009 | 9.000 | 23.205 | 1.623E-03 | 0.966 | 60.676 | 4.029E-04 | 8.650E-03 |
| | NN-2layer | 0.140 | 3.292 | 12.381 | 2.251E-04 | 0.259 | 28.179 | 2.416E-04 | 6.740E-03 |
| | NN-3layer | 0.165 | 3.442 | 13.921 | 2.675E-04 | 0.441 | 21.634 | 2.159E-04 | 7.439E-03 |
| | RF | 0.557 | 27.237 | 20.469 | 1.156E-03 | 2.910 | 41.068 | 3.170E-04 | 8.813E-03 |
| | XGB | 0.753 | 13.107 | 21.086 | 1.225E-03 | 1.401 | 40.828 | 3.264E-04 | 7.688E-03 |
| Max Error | LUT | 1.391 | 71.779 | 132.727 | 4.490E-03 | 7.533 | 105.551 | 2.138E-02 | 5.251E-02 |
| | PLS | 9.103 | 269.837 | 290.055 | 5.381E-03 | 28.877 | 867.291 | 1.890E-03 | 2.520E-02 |
| | SVR | 11.965 | 31.809 | 78.506 | 8.506E-03 | 2.636 | 592.881 | 2.714E-03 | 2.471E-02 |
| | GP | 3.736 | 48.060 | 84.890 | 5.491E-03 | 3.635 | 587.723 | 2.797E-03 | 2.414E-02 |
| | NN-2layer | 8.158 | 257.293 | 272.003 | 6.019E-03 | 27.727 | 851.586 | 3.658E-03 | 3.658E-02 |
| | NN-3layer | 8.750 | 255.192 | 271.572 | 6.123E-03 | 27.313 | 840.176 | 3.323E-03 | 4.400E-02 |
| | RF | 4.367 | 61.402 | 98.132 | 6.079E-03 | 6.801 | 419.489 | 2.675E-03 | 3.559E-02 |
| | XGB | 4.122 | 52.837 | 92.469 | 8.091E-03 | 5.629 | 389.348 | 2.813E-03 | 2.561E-02 |
| Min Error | LUT | -0.726 | -80.039 | -65.362 | -3.337E-04 | -8.379 | -53.778 | -3.048E-05 | -4.025E-02 |
| | PLS | -9.631 | -247.385 | -358.316 | -8.859E-03 | -26.308 | -673.366 | -5.995E-03 | -7.506E-02 |
| | SVR | -3.623 | -83.015 | -110.200 | -4.037E-03 | -8.817 | -148.101 | -2.211E-03 | -5.306E-02 |
| | GP | -1.923 | -54.736 | -60.597 | -1.516E-03 | -6.649 | -200.431 | -1.799E-03 | -5.503E-02 |
| | NN-2layer | -7.603 | -261.059 | -277.970 | -4.779E-03 | -27.745 | -713.290 | -3.503E-03 | -7.924E-02 |
| | NN-3layer | -7.711 | -268.349 | -261.718 | -4.828E-03 | -27.204 | -785.691 | -3.463E-03 | -7.766E-02 |
| | RF | -2.087 | -64.946 | -67.003 | -3.154E-03 | -6.844 | -154.989 | -1.748E-03 | -5.507E-02 |
| | XGB | 1.947 | -56.110 | -92.003 | -4.932E-03 | -5.485 | -225.811 | -2.019E-03 | -5.460E-02 |

Table C.2. Predictive performance of LUT vs. ML algorithms for 3-cycle rods (green denotes the best desirability, and blue indicates the worst performance)

| | | Max Plenum Pressure [MPa] | Max Hydrogen Concentration [ppm] | Max Hoop Stress [MPa] | Max Hoop Strain [-] | Max Oxide Thickness [μm] | Max Fuel Temperature [K] | PCI-SCC Failure Probability [-] | PCI-MPS Failure Probability [-] |
|---|---|---|---|---|---|---|---|---|---|
| $R^2$ | LUT | 0.960 | 0.445 | -5.742 | -19.417 | 0.440 | 0.912 | -858.062 | -2.111 |
| | PLS | 0.016 | 0.893 | 0.451 | -22.284 | 0.893 | -1.336 | -0.178 | -0.063 |
| | SVR | 0.017 | 0.952 | 0.453 | -14.403 | 0.951 | -0.824 | 0.475 | -0.368 |
| | GP | 0.348 | 0.920 | 0.535 | -16.085 | 0.920 | -0.461 | 0.050 | 0.006 |
| | NN-2layer | 0.977 | 0.986 | 0.898 | 0.668 | 0.983 | 0.784 | 0.050 | -0.104 |
| | NN-3layer | 0.983 | 0.985 | 0.947 | 0.631 | 0.980 | 0.909 | 0.849 | 0.051 |
| | RF | 0.592 | 0.483 | 0.693 | -7.465 | 0.476 | 0.319 | 0.056 | -0.146 |
| | XGB | 0.469 | 0.780 | 0.619 | -11.063 | 0.786 | 0.335 | 0.071 | -0.078 |
| RMSE | LUT | 0.292 | 30.957 | 84.742 | 2.612E-03 | 3.248 | 13.549 | 9.525E-03 | 1.728E-02 |
| | PLS | 1.456 | 13.614 | 24.179 | 2.789E-03 | 1.433 | 69.976 | 3.528E-04 | 1.010E-02 |
| | SVR | 1.455 | 9.152 | 24.132 | 2.268E-03 | 0.972 | 61.833 | 2.355E-04 | 1.146E-02 |
| | GP | 1.185 | 11.766 | 22.244 | 2.389E-03 | 1.238 | 55.331 | 3.167E-04 | 9.769E-03 |
| | NN-2layer | 0.222 | 4.868 | 10.425 | 3.331E-04 | 0.576 | 21.267 | 3.167E-04 | 1.029E-02 |
| | NN-3layer | 0.192 | 5.051 | 7.489 | 3.511E-04 | 0.626 | 13.832 | 1.262E-04 | 9.545E-03 |
| | RF | 0.937 | 29.884 | 18.076 | 1.682E-03 | 3.170 | 37.790 | 3.158E-04 | 1.049E-02 |
| | XGB | 1.069 | 19.504 | 20.156 | 2.007E-03 | 2.026 | 37.346 | 3.132E-04 | 1.017E-02 |
| MAE | LUT | 0.243 | 23.109 | 80.012 | 2.224E-03 | 2.427 | 10.994 | 8.152E-03 | 1.523E-02 |
| | PLS | 1.378 | 10.450 | 20.721 | 2.714E-03 | 1.099 | 64.011 | 3.067E-04 | 7.181E-03 |
| | SVR | 1.120 | 7.293 | 19.340 | 1.923E-03 | 0.780 | 44.810 | 1.750E-04 | 7.847E-03 |
| | GP | 0.956 | 9.315 | 18.051 | 2.139E-03 | 0.979 | 46.973 | 2.268E-04 | 6.367E-03 |
| | NN-2layer | 0.183 | 3.667 | 8.656 | 2.483E-04 | 0.401 | 17.350 | 2.378E-04 | 7.358E-03 |
| | NN-3layer | 0.170 | 3.743 | 5.962 | 2.819E-04 | 0.488 | 10.432 | 7.795E-05 | 6.357E-03 |
| | RF | 0.649 | 26.964 | 14.268 | 1.317E-03 | 2.853 | 27.819 | 2.132E-04 | 6.770E-03 |
| | XGB | 0.788 | 16.460 | 16.158 | 1.682E-03 | 1.719 | 30.538 | 2.174E-04 | 6.553E-03 |
| Max Error | LUT | 0.298 | 27.503 | 132.727 | 4.413E-03 | 2.888 | 43.769 | 1.896E-02 | 4.259E-02 |
| | PLS | 5.945 | 188.753 | 189.996 | 5.427E-03 | 19.949 | 290.867 | 1.021E-03 | 2.145E-02 |
| | SVR | 6.417 | 14.406 | 86.206 | 1.013E-02 | 1.524 | 560.762 | 7.368E-04 | 2.330E-02 |
| | GP | 3.807 | 17.272 | 62.667 | 6.535E-03 | 1.851 | 134.283 | 1.622E-03 | 1.632E-02 |
| | NN-2layer | 3.981 | 207.066 | 190.621 | 3.547E-03 | 20.893 | 219.602 | 1.833E-03 | 2.914E-02 |
| | NN-3layer | 4.097 | 203.698 | 184.866 | 3.458E-03 | 21.435 | 230.491 | 1.789E-03 | 3.952E-02 |
| | RF | 6.798 | 35.385 | 57.281 | 5.102E-03 | 3.501 | 174.399 | 1.201E-03 | 2.145E-02 |
| | XGB | 4.904 | 23.289 | 58.739 | 6.539E-03 | 3.872 | 153.388 | 1.323E-03 | 1.914E-02 |
| Min Error | LUT | -0.623 | -80.039 | 0.954 | -2.460E-04 | -8.379 | -30.895 | 8.637E-07 | -3.029E-02 |
| | PLS | -4.081 | -211.993 | -173.074 | -2.112E-03 | -22.320 | -180.281 | -2.720E-03 | -6.184E-02 |
| | SVR | -5.419 | -34.841 | -71.869 | -2.764E-03 | -4.535 | -185.503 | -1.133E-03 | -5.149E-02 |
| | GP | -1.174 | -43.148 | -48.886 | -1.059E-03 | -4.573 | -71.692 | -1.079E-03 | -5.370E-02 |
| | NN-2layer | -3.874 | -202.169 | -187.594 | -2.731E-03 | -22.477 | -276.825 | -1.843E-03 | -7.179E-02 |
| | NN-3layer | -3.939 | -209.653 | -181.088 | -2.712E-03 | -22.659 | -237.486 | -1.795E-03 | -7.172E-02 |
| | RF | -0.686 | -64.495 | -53.070 | -1.502E-03 | -6.794 | -44.335 | -1.157E-03 | -5.937E-02 |
| | XGB | -0.984 | -58.236 | -57.503 | -8.137E-04 | -5.907 | -41.503 | -9.915E-04 | -5.605E-02 |

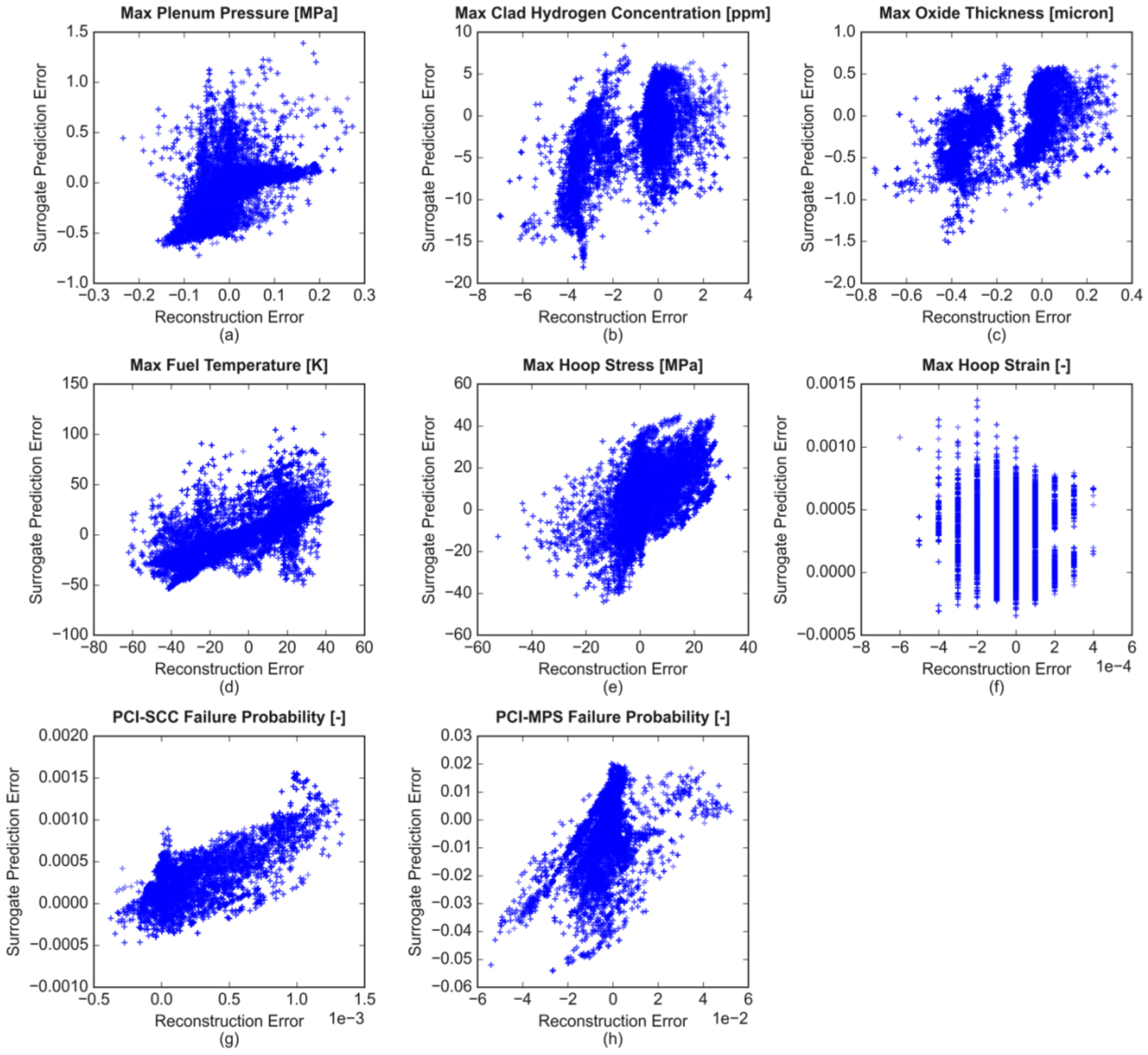

Figure C.2. Surrogate prediction error versus reconstruction error with the selected best surrogate for the eight fuel performance QoIs for 2-cycle rods.

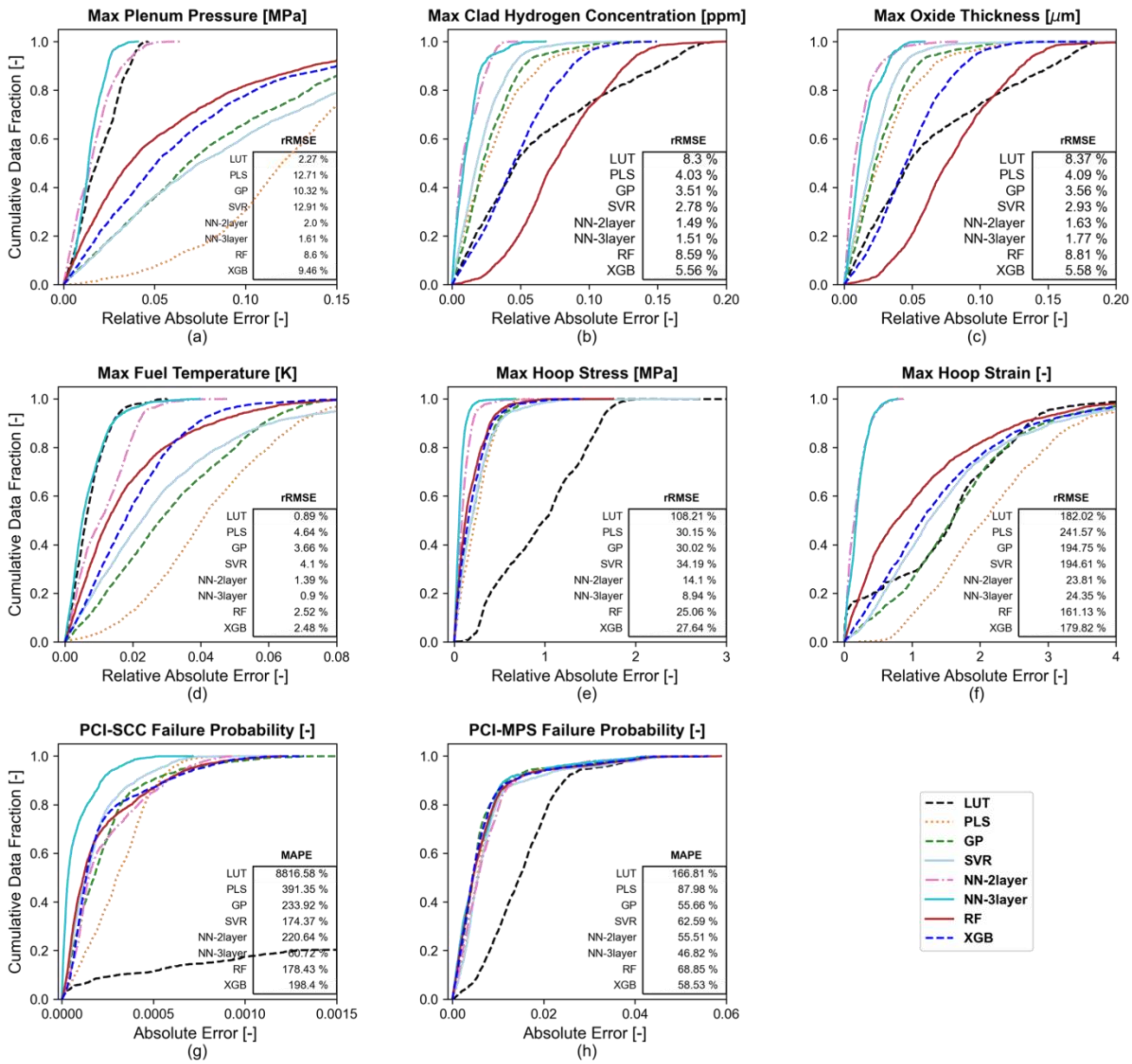

Figure C.3. Cumulative data fraction of (relative) absolute errors for LUT and ML algorithms for the eight fuel performance QoIs for 3-cycle rods.

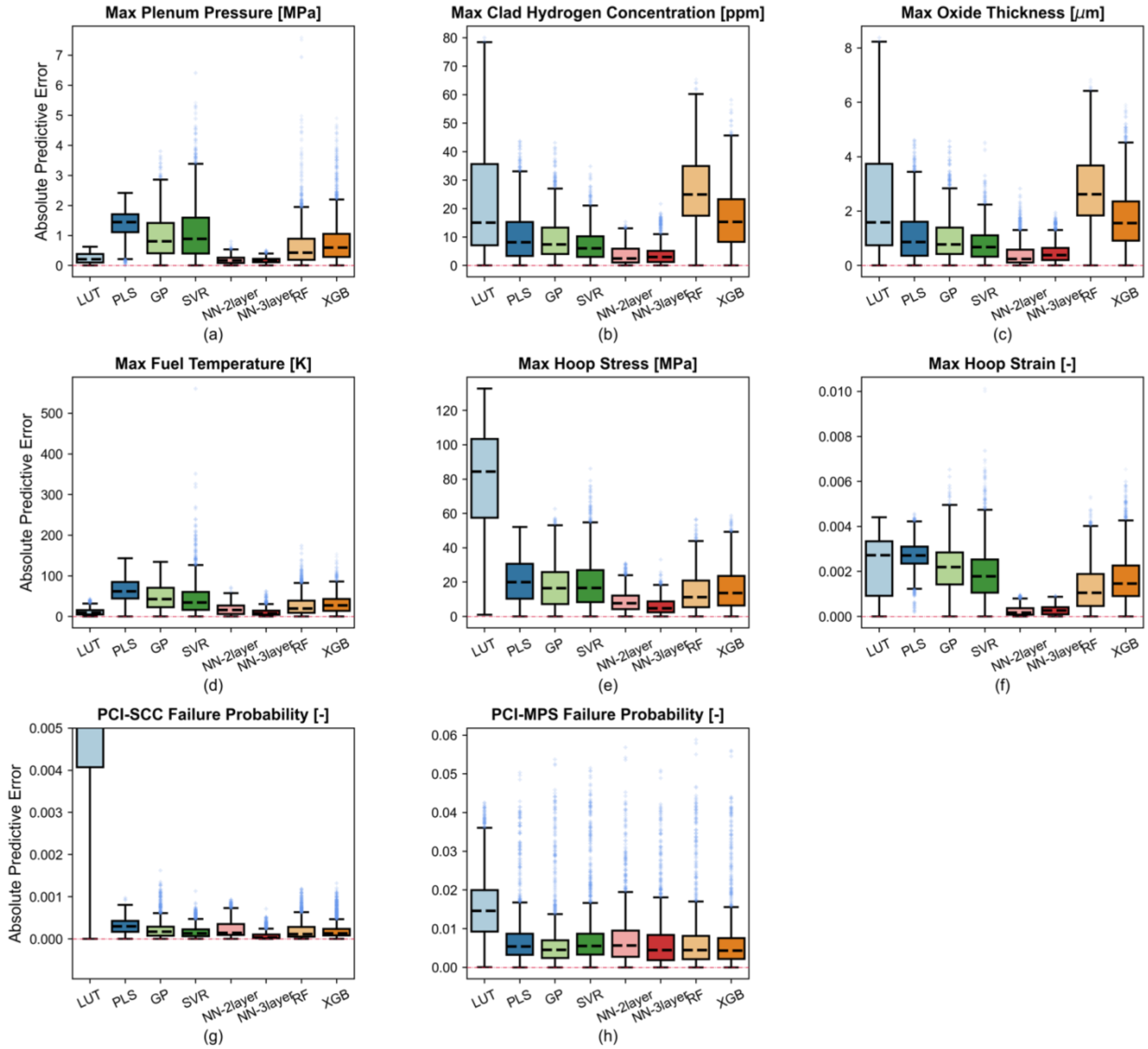

Figure C.4. Boxplots of absolute predictive error with LUT and ML algorithms for the eight fuel performance QoIs for 3-cycle rods.